\documentclass[twocolumn]{aastex63}
\usepackage{enumitem}
\shorttitle{An Observation Strategy for the Target Stars for CHES}
\shortauthors{Tan et al.}

\graphicspath{{./}{figures/}}

\begin{document}

\title{Closeby Habitable Exoplanet Survey (CHES). II. An Observation Strategy for the Target Stars}

\author{Dongjie Tan}
\affiliation{CAS Key Laboratory of Planetary Sciences, Purple Mountain Observatory, Chinese Academy of Sciences, Nanjing 210023, China;jijh@pmo.ac.cn}
\affiliation{School of Astronomy and Space Science, University of Science and Technology of China, Hefei 230026, China}

\author{Jianghui Ji}
\affiliation{CAS Key Laboratory of Planetary Sciences, Purple Mountain Observatory, Chinese Academy of Sciences, Nanjing 210023, China;jijh@pmo.ac.cn}
\affiliation{School of Astronomy and Space Science, University of Science and Technology of China, Hefei 230026, China}
\affiliation{CAS Center for Excellence in Comparative Planetology, Hefei 230026, China}

\author{Chunhui Bao}
\affiliation{CAS Key Laboratory of Planetary Sciences, Purple Mountain Observatory, Chinese Academy of Sciences, Nanjing 210023, China;jijh@pmo.ac.cn}
\affiliation{School of Astronomy and Space Science, University of Science and Technology of China, Hefei 230026, China}

\author{Xiumin Huang}
\affiliation{CAS Key Laboratory of Planetary Sciences, Purple Mountain Observatory, Chinese Academy of Sciences, Nanjing 210023, China;jijh@pmo.ac.cn}
\affiliation{School of Astronomy and Space Science, University of Science and Technology of China, Hefei 230026, China}

\author{Guo Chen}
\affiliation{CAS Key Laboratory of Planetary Sciences, Purple Mountain Observatory, Chinese Academy of Sciences, Nanjing 210023, China;jijh@pmo.ac.cn}
\affiliation{School of Astronomy and Space Science, University of Science and Technology of China, Hefei 230026, China}

\author{Su Wang}
\affiliation{CAS Key Laboratory of Planetary Sciences, Purple Mountain Observatory, Chinese Academy of Sciences, Nanjing 210023, China;jijh@pmo.ac.cn}
\affiliation{CAS Center for Excellence in Comparative Planetology, Hefei 230026, China}

\author{Yao Dong}
\affiliation{CAS Key Laboratory of Planetary Sciences, Purple Mountain Observatory, Chinese Academy of Sciences, Nanjing 210023, China;jijh@pmo.ac.cn}
\affiliation{CAS Center for Excellence in Comparative Planetology, Hefei 230026, China}

\author{Haitao Li}
\affiliation{National Space Science Center, Chinese Academy of Sciences, Beijing 100190, China}
\affiliation{University of Chinese Academy of Sciences, Beijing 100049, China}

\author{Junbo Zhang}
\affiliation{Institute of Optics and Electronics, Chinese Academy of Sciences, Chengdu 610209, China}
\affiliation{Key Laboratory on Adaptive Optics, Chinese Academy of Sciences, Chengdu 610209, China}

\author{Liang Fang}
\affiliation{University of Chinese Academy of Sciences, Beijing 100049, China}
\affiliation{Institute of Optics and Electronics, Chinese Academy of Sciences, Chengdu 610209, China}

\author{Dong Li}
\affiliation{Innovation Academy for Microsatellites of Chinese Academy of Sciences, Shanghai 201306, China}

\author{Lei Deng}
\affiliation{Innovation Academy for Microsatellites of Chinese Academy of Sciences, Shanghai 201306, China}

\author{Jiacheng Liu}
\affiliation{School of Astronomy and Space Science, Nanjing University, Nanjing 210046, China}

\author{Zi Zhu}
\affiliation{School of Astronomy and Space Science, Nanjing University, Nanjing 210046, China}
\affiliation{University of Chinese Academy of Sciences, Nanjing 211135, China}

\begin{abstract}

The Closeby Habitable Exoplanet Survey (CHES) constitutes a mission intricately designed to systematically survey approximately 100 solar-type stars located within the immediate proximity of the solar system, specifically within a range of 10 parsecs. The core objective of this mission is the detection and characterization of potentially habitable Earth-like planets or super-Earths within the habitable zone of these stars. The CHES mission obtains high-precision astrometric measurements of planets orbiting the target stars by observing angular distance variations between the target star and reference stars. As a result, we surveyed the relevant parameters of both target and reference stars in detail, conducting a thorough analysis and calculation of the required observation accuracy, the number of observations, and the priority assigned to each target star.
Observational emphasis will be concentrated on targets considered of higher priority, ensuring the effectiveness of their observation capabilities. Through this approach, we formulate a five-year observation strategy that will cover all the target stars within a six-month timeframe. The strategy not only fulfills the required observing capability but also exhibit high efficiency simultaneously, providing an executable program for future mission. Over the span of the mission's five-year duration, a cumulative observation time of 29,220 hours will be available. Approximately 86 percent of this, totaling 25,120 hours, is allocated for the observation of target stars. This allocation leaves approximately 4,100 hours for extended scientific observation programs. We have also performed simulated observations based on this strategy and verified its observational capability for exoplanets.

\end{abstract}

\keywords{Astrometric exoplanet detection (2130) --- High angular resolution (2167) --- Telescope focal plane photography (1687)}

\section{Introduction} \label{sec:intro}
The exploration of habitable planets bears significance in understanding the formation of our home planet, its future evolution, and addressing the existential question of our solitude in the universe. Advances in observational methodologies and techniques have steadily bolstered the detection of exoplanets, marking a discernible trajectory of refinement in observational precision. The first exoplanet of 51 Pegasi b, was discovered to orbit a sun-like star \citep{Mayor1995} using radial velocity method, which revolutionizes our understanding of planetary systems beyond our own. Subsequently, high-precision photometric measurements were conducted on the star HD 209458, revealing a gas giant transiting across the stellar disk \citep{Charbonneau2000}. With advancements in high-contrast and high-angular resolution instruments, direct imaging for the planet around 2M1207 were also conducted using infrared photometry techniques \citep{Chauvin2004}. The astrometric method has also contributed to the detection of exoplanets \citep{Sahlmann2013, Curiel2022}. {One of them is an exoplanet discovered by Gaia \citep{Sozzetti2023}. By analyzing the astrometric perturbations in the Gaia DR3, the semi-major axis of the host star's orbit can be determined, which allows for the inference of the planet's orbit and mass. This finding was further confirmed and refined through radial velocity (RV) observations, which provided corrections and narrowed down the actual range of the planet's parameters.} Currently, more than 6000 exoplanets are discovered, which unveils a diversity and complexity of the planetary systems that include hot Jupiters, cold-Jupiters, mini-Neptunes, super-Earths, and terrestrial planets around the stars \citep{Butler2004,Borucki2010,Borucki2011,Batalha2013,Anglada2016,Gillon2017}.{ Most of the small planets in temperate orbits have been found around M dwarfs due to the much higher sensitivity of the transit and Doppler spectroscopy technique on the low-mass stars. As a result, there is a remarkable absence of detected habitable terrestrial planets orbiting solar-type stars among the reported population.}

{In recent years, a number of programs have been conducted to infer the presence of potentially habitable planets.} The Kepler mission and its successor, the K2 mission \citep{Howell2014}, employ the transiting method for planetary detection. This method relies on high-precision photometry and the utilization of a large field of view (FOV) to simultaneously monitor multiple targets. Its main objective is to identify small-mass, short-period planets orbiting dwarf stars, with K2 having already confirmed the discovery of 548 planets. The TESS mission \citep{Ricker2015} is dedicated to the search for planets transiting nearby bright stars, observing stars one to two orders of magnitude brighter than those scrutinized in the Kepler mission. Over its nearly six-year operational span, TESS has confirmed 410 planets. {Given that the transiting method is inclined to detecting planets with relatively short orbital periods, most observed orbital periods do not exceed 10 days.} Furthermore, the James Webb Space Telescope (JWST) is capable of characterizing low-mass planets orbiting M dwarfs. The seven planets of TRAPPIST-1 \citep{Gillon2017} present excellent targets for JWST, allowing for the detection of planetary atmospheres using the transiting method \citep{Lustig-Yaeger2019}. In addition to space telescopes, large ground-based telescopes can also play a crucial role in planetary detection. Techniques such as transmission spectroscopy \citep{Snellen2013,Rodler2014} or direct imaging during transits \citep{Crossfield2013} are employed to study atmospheric components and other biological features, via the Thirty Meter Telescope (TMT) \citep{Skidmore2015}, the Giant Magellan Telescope (GMT) \citep{Johns2012S}, and the European Extremely Large Telescope (E-ELT) \citep{Gilmozzi2007}.

In the future, additional missions are expected to actively target the detection and characterization of exoplanets through a variety of techniques. The PLATO mission \citep{Rauer2014}, scheduled for launch in 2026, is designed to discern the radii of planets as they transit across stars while simultaneously investigating their atmospheric components. Atmospheric Remote-sensing Infrared mission Exoplanet Large-survey (ARIEL) \citep{Tinetti2016} mission aims to study the composition, formation, and evolution by surveying hundreds of transiting exoplanets in both visible and infrared wavelengths. Similar to SIM \citep{Sozzetti2002, Sozzetti2003, Catanzarite2006}, the JASMINE mission \citep{Kawata2023} is proposed to be an astrometric satellite by observing the exoplanets at the near-infrared wavelengths \citep{Baba2020}.  The JASMINE satellite is scheduled for launch in 2028, with positional measurements of stars expected to be accurate to the order of tens of micro-arcseconds. Morever, the HabEX mission \citep{Gaudi2020} is dedicated to detecting planetary atmospheres through a space-based direct imaging method. Furthermore, Habitable Worlds Observatory (HWO) survey \citep{Mamajek2024} is anticipated to be launched in the 2040s. The Large Interferometer For Exoplanets (LIFE) mission \citep{Quanz2022} and Miyin Project \citep{Ji2020,Wang2024},  aim to detect nearby habitable planets through space interferometry.

Unlike the detection methods of transit and the radial velocity, the CHES mission \citep{Ji2022,Ji2024,Bao2024A} will employ an astrometric technique to discover habitable-zone planets around nearby solar-type stars and is a candidate for a future space mission by the Chinese Academy of Sciences. The signals derived from this motion provide valuable information, including the period of the planet's motion and its mass ratio. To illustrate the precision required for detection of terrestrial planets, consider a solar-type star situated 10 pc away, harboring a $M_{\oplus}$  planet and orbiting at 1 au away from the star. The allowable measurement errors for position, enabling the detection of the planet's existence, is ${0.3~ \mu \rm as}$ \citep{Malbet2012}. As the currently highest-precision star catalog, Gaia DR3 can also detect cold Jupiter-mass planets over its observations \citep{Lattanzi2000, Sozzetti2001, Casertano2008}. However, the precision of its measurements for reference stars in the magnitude range of 9-12, including positions, proper motions, and parallaxes, exceeds 10 microarcseconds \citep{Lindegren2021}, which may not be sufficient for establishing a reference system with the required accuracy for space observations. Consequently, they cannot detect perturbation signals on the order of micro-arcsecond produced by Earth-like planets.

CHES will utilize relative measurements to attain astrometric parameters for neighboring solar-type stars with accuracies on the order of micro-arcseconds. The planetary data is inferred from the micro-arcsecond-scale angular distance variation between the target star and the reference star. In contrast to a sky survey telescope \citep{vanLeeuwen2007a,GaiaCollaboration2016}, CHES deliberately focuses on the observation of nearly 100 pre-selected solar-type target stars. {For these target stars, CHES will conduct extensive observations throughout its operational period to achieve an observational precision better than Gaia (${20~ \mu \rm as}$ )\citep{Lindegren2021}, reaching the approximately ${1~ \mu \rm as}$ required for detecting habitable planets.} In addition to ensuring precise observations of target stars, it is crucial to distribute the observations as evenly as possible across the five years of operation. To maintain observing efficiency, it is imperative to minimize the telescope's back-and-forth movements over large angles during pointing.

This study aims to compute the necessary observing accuracy for each target star, considering the relevant parameters of both the target and reference stars. Through this approach, our objective is to formulate an observing strategy that not only meets the required observing capability but also exhibits high efficiency. We also validate the efficacy of this strategy for planetary detection through simulations.

This work is structured as follows: In Section~\ref{sect:mission}, we present an overview of the CHES mission, analyzing the expected observed values resulting from the observations. Section~\ref{sect:Photon} explores the determination of the number of observations and the optimal single exposure time for each target star considering photon noise. The prioritization of observations for target stars is addressed in Section~\ref{sect:Prior}. The observing strategy is given in Section~\ref{sect:Strategy} while Section~\ref{sect:Impact} assesses the impact of observing strategy on the mission. Finally, we provide a brief summary.

\section{The CHES mission and observed quantity} \label{sect:mission}
\begin{figure*}
	\plotone{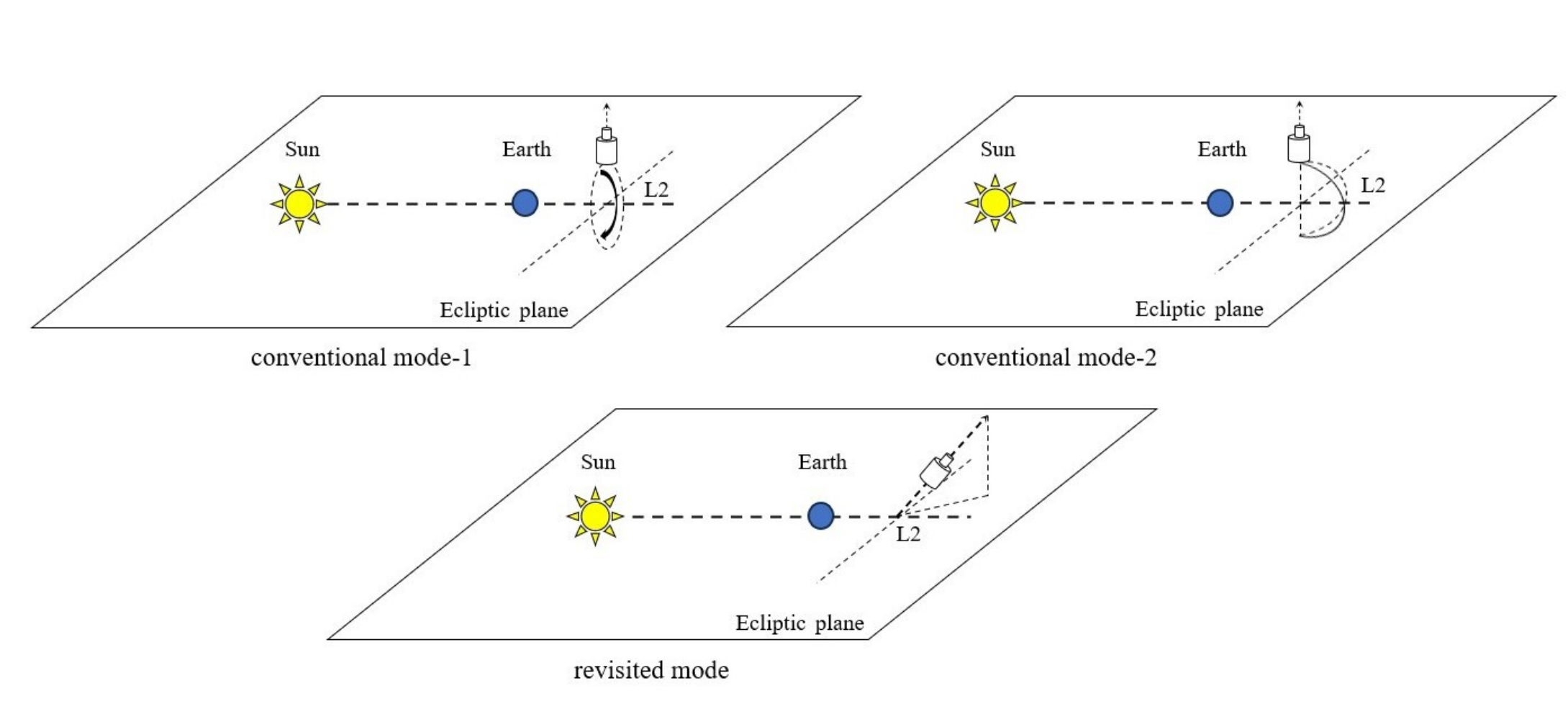}
	\caption{Two types of the observation modes \citep{Ji2022}. \textit{Upper panel}: conventional mode. \textit{Lower panel}: revisited mode.}
	\label{mode}
\end{figure*}

The satellite's in-orbit operation encompasses two observation modes: conventional mode and revisited mode, as illustrated in Figure ~\ref{mode}. In the conventional mode, two types are distinguished. The first involves the rotation of the telescope's optical axis along the great circle passing through the north and south ecliptic poles. The second entails the movement of the telescope's optical axis along two semicircles connected at the north and south ecliptic poles, forming an angle of 60°, observing nearby target stars. In the revisited mode, there are no restrictions on the optical axis pointing of the telescope, except for avoiding direct sunlight. This mode allows the telescope to be pointed at any target star for observation. A more comprehensive description of these modes and the associated observing strategy will be presented in Section ~\ref{sect:Strategy}. In conjunction with the two scanning modes, sufficient observations of all target stars will be conducted efficiently, as will be discussed later.

The relative measurement method employed differs fundamentally from the absolute measurement method \citep{Kovalevsky2004}, which relies on the precision of a prior catalog. The accuracy of the relative measurement hinges on the precise measurement of angular distances. In absolute measurement, obtaining the positions of reference stars from a prior catalog is imperative, followed by the establishment of a reference frame based on these positions. Subsequently, the positional change information of the target star is ascertained within this reference frame on the {local coordinate frame around each target star}. Consequently, the accuracy of target star position measurement is limited by the precision of the prior catalog. Even with the most accurate astrometric data available to date, such as Gaia DR3 \citep{GaiaCollaboration2021}, with precision in the order of tens of micro-arcseconds for position, it remains insufficient for detecting stellar disturbance signals caused by Earth-like planets.

The relative measurement method involves measuring the angular distance between the target star and the reference star within the {local reference frame spanning the telescope field of view}. This change arises from the disparity in proper motion and parallax between the two stars, coupled with the perturbation of the planets affecting the position of the target star. {The parameters of relative measurement are illustratively described in Eq.~\ref{l},}
\begin{equation}
	\label{l}
	\begin{array}{cc}
	l_{\alpha} (t)=\Delta \alpha_{0}+\Delta \mu_{\alpha}(t-t_{0})+(\pi _{1}F_{\alpha1}-\pi_{2}F_{\alpha2})\\
	+ ~\alpha_{p}(t) + O(t^{2}) \\
	l_{\delta} (t)=\Delta \delta_{0}+\Delta \mu_{\delta }(t-t_{0})+(\pi _{1}F_{\delta1}-\pi_{2}F_{\delta2})\\
	+ ~\delta_{p}(t) + O(t^{2})
	\end{array}
\end{equation}
{where $l_{\alpha}$ and $l_{\delta}$ represent two components of the angular distance. $\Delta \alpha_{0}$ and $\Delta \delta_{0}$ denote the initial coordinate difference, while $\Delta \mu_{\alpha}$ and $\Delta \mu_{\delta}$ stand for the proper motion difference. The $F_{\alpha1},F_{\alpha2},F_{\delta1},F_{\delta1}$ represent the components of the parallax factor of two stars in the directions of right ascension and declination, respectively, and $\alpha_{p}$ and $\delta_{p}$ account for the variations in position due to the planets, while $O(t^{2})$ represents the higher-order terms of the proper motion.}

In Eq.~\ref{l}, there is no necessity to ascertain the actual position coordinates of the reference stars, eliminating the requirements for a prior catalog. Consequently, the measurement accuracy of angular distance has become a crucial factor in determining the ability of exoplanet detection. {To achieve such high-precision astrometry, Theia previously imposed strict requirements on the instruments \citep{Malbet2021} and attempted to calibrate telescope distortions by using the positions of reference stars obtained from the Gaia prior star catalog \citep{Malbet2022}. CHES also considered the low-distortion wide-field telescope optical systems, high-stability attitude control and high-precision thermal control technologies to ensure measurements reach micro-arcsecond accuracy in angular distance \citep{Ji2022,Ji2024}.}

\begin{figure*}
	\gridline{\fig{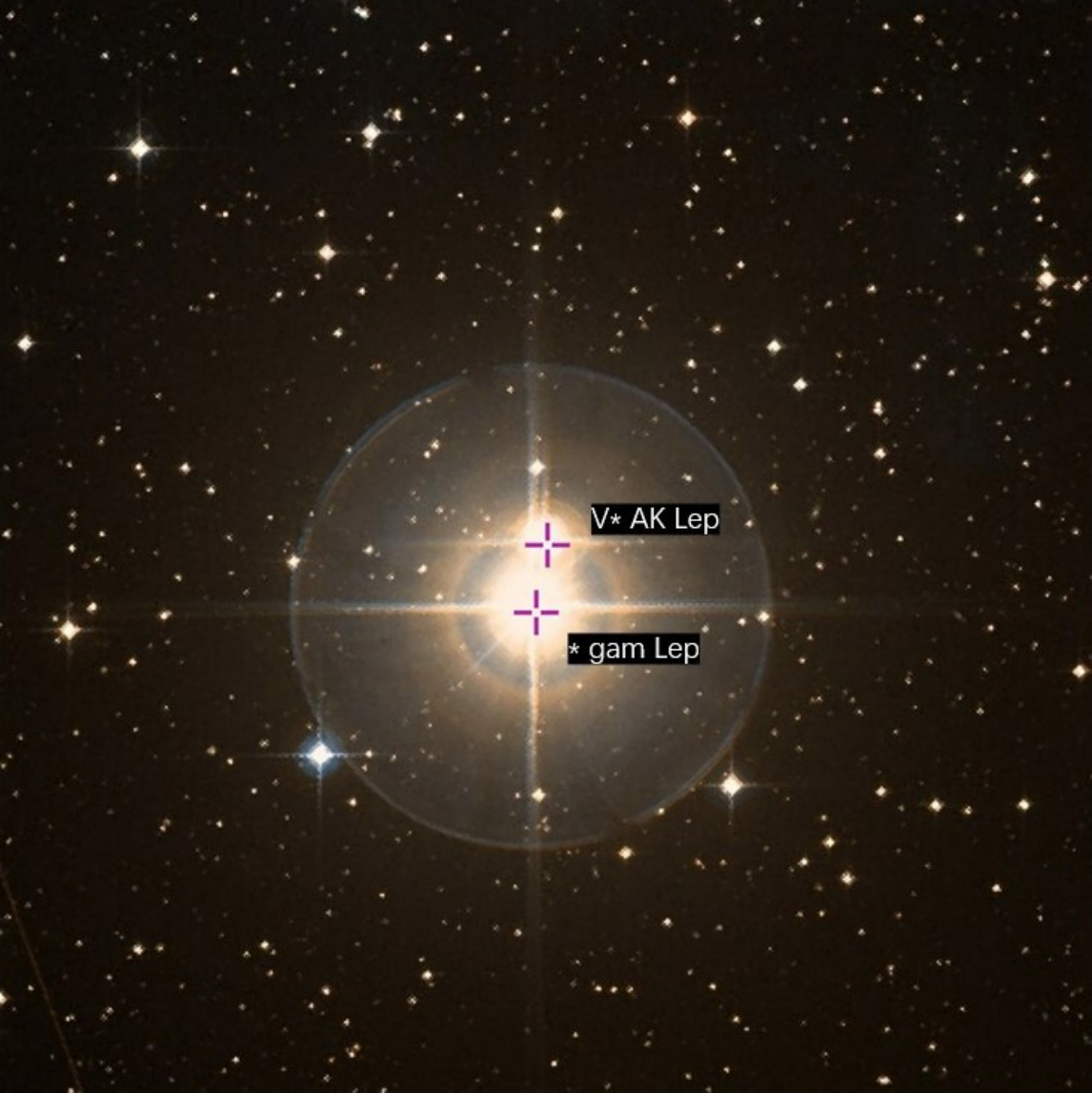}{0.4\textwidth}{(a)}
		\fig{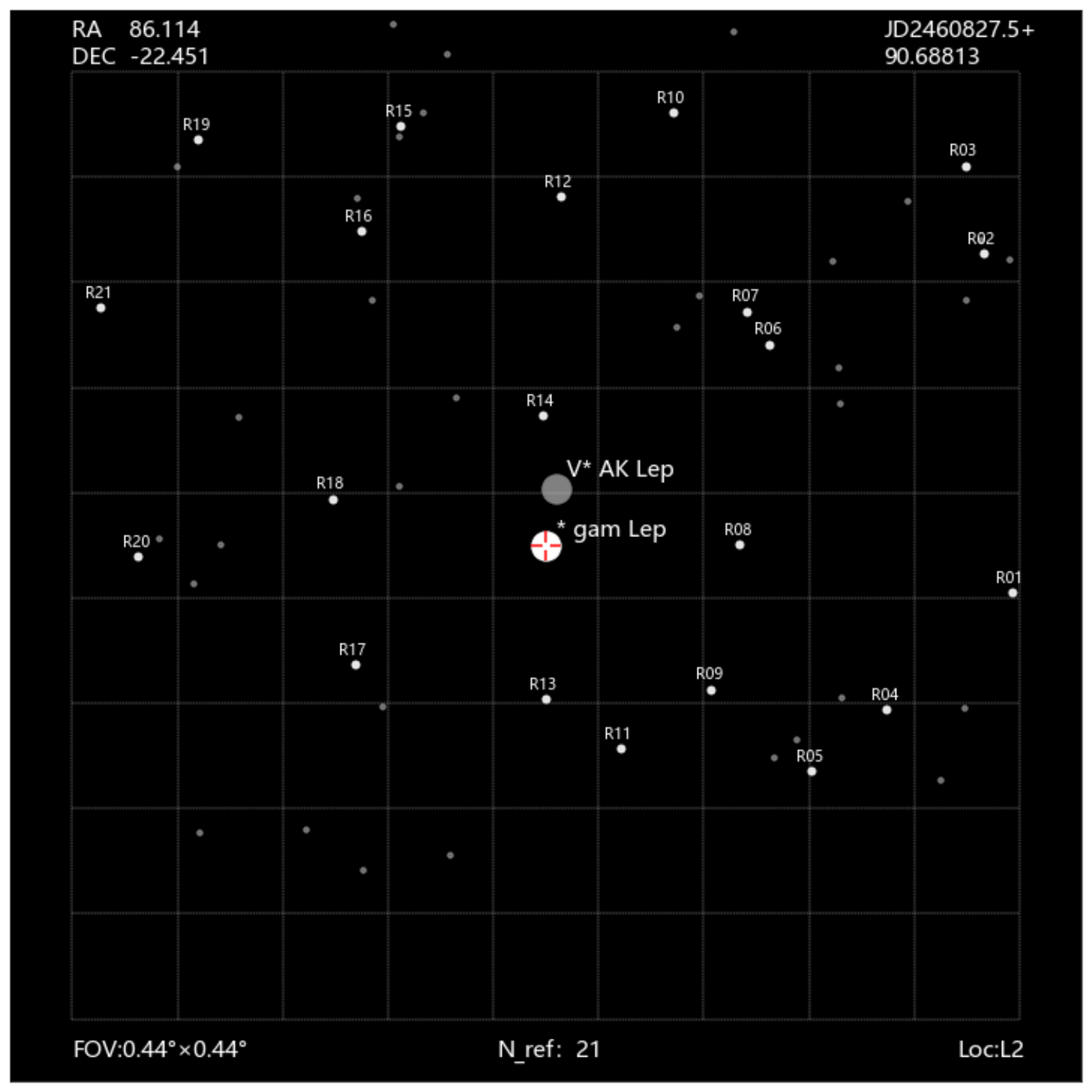}{0.4\textwidth}{(b)}
	}
	\caption{The relative positions of V* AK Lep and * gam Lep on the celestial sphere. While these two stars do not form a binary star system, their close proximity on the celestial sphere allows for simultaneous observation of both targets. The FOV is $0.44^{\circ} \times 0.44^{\circ}$. \emph{Panel (a)}: Image from Aladin Sky Atlas \citep{Bonnarel2000}. \emph{Panel (b)}: Simulated observational images of the target stars \citep{Ji2023}.}
	\label{gam Lep}
\end{figure*}

The measurement of angular distance in actual observations involves determining the position and distance between the target star and the reference star on the focal plane. It is crucial to consider the projection relationship from the celestial sphere to the focal plane when addressing angular distance on the celestial sphere and distance on the focal plane. During the observation process, no reference frame is available in the FOV as it does not rely on a priori catalog. Typically, the target star is positioned at the center of the FOV, surrounded by multiple distributed reference stars. Special cases may arise, such as when the target of observation is a binary star, exemplified by * i Boo \citep{vanLeeuwen2007b}, where eleven target stars fall into this category. In such cases, the center of the FOV is directed toward the brighter of the two stars, and the reference stars for both targets within the FOV are identical. Consequently, the two targets can be observed simultaneously, enhancing observation efficiency. An even more uncommon scenario is illustrated by the stars V* AK Lep and * gam Lep \citep{Cutri2003}, which are not binaries but are positioned so closely on the celestial sphere that they can be observed concurrently. Figure ~\ref{gam Lep} displays the relative positions of these two stars on the celestial sphere.

In previous study, the direction of the optic axis was a subject of discussion \citep{Tan2022}. This discussion highlighted that the target star may not be precisely at the center of the FOV. However, it was established that as long as the range of the point of change falls within a specific range (68 milli-arcseconds at 1 micro-arcsecond), it satisfies the error requirements.

In the CHES mission, the FOV is $0.44^{\circ} \times 0.44^{\circ}$, and the focal plane is equipped with an array of 9 by 9 scientific CMOS detectors. Each detector comprises $4096 \times 4096$ pixels. Utilizing the projection relationship and the pertinent instrument parameters, a nonlinear relationship between the radians value of angular distance and the corresponding pixel value is established. One end of the angular distance is anchored at the center of the FOV:
{\begin{equation}
	P_{D}=\frac{P_{N}}{\tan R} \tan l,
\end{equation}
where $P_{D}$ represents the pixel distance from the center of the image point to the center of the FOV. $P_{N}$ refers to the number of pixels from the center to the edge of the detector, with a value of $9\times4096/2$ in this case. $R$ represents the angular radius from the center to the edge of the FOV, as $0.44^{\circ}/2$.} $l$ denotes the angular distance between the stars on the celestial sphere and the center of the FOV. The relationship between the pixel distance from the center point and the angular distance from the center point on the celestial sphere is not strictly linear but exhibits an approximate linearity.

The angular distance between the target star and the reference star can also be employed for the indirect measurement of changes in angular distance between reference stars, utilizing spherical triangulation. These changes in angular distance among reference stars are also on the order of micro-arcseconds, providing valuable information for scientific inquiries such as identifying the presence of binaries in reference stars.

\section{Photon noise and observation time}
\label{sect:Photon}
In observations at the micro-arcsecond scale, the impact of photon noise on observational accuracy becomes non-negligible. The error due to photon noise is inversely proportional to the number of photons received, with lower noise levels for a target star surrounded by a greater number of reference stars in the FOV. Consequently, no more than 12 brightest reference stars, along with the target star, are chosen to compute the number of photons received per unit time. The incorporation of additional reference stars increases the photon count for them, thereby reducing the influence of photon noise.

The error arising from photon noise is directly proportional to the square root of the inverse of the product of the number of observations and the exposure time for each observation, expressed as $\sigma \propto (N_{vis} \times t_{vis})^{-1/2}$ \citep{Malbet2021}. Consequently, increasing both the number of observations and the exposure time for each observation proves effective in reducing photon noise.  We hope to discover Earth-like planets around the target star, and therefore need to observe signals of periodic motions of the host star due to the presence of Earth-like planets in the habitable zone of the target star.

We calculate the range of habitable zone of the Earth-like planets based on the effective temperature and luminosity of each target star \citep{Kopparapu2014}:
\begin{equation}
	S_{eff}=S_{eff\odot }+aT_{\star }+bT_{\star}^{2}+cT_{\star}^{3}+dT_{\star}^4,
\end{equation}
where $T_{\star}=T_{eff} - 5780~\rm{K}$ and the coefficients of $T_{\star}$ are parameters related to the planetary mass. In this context, the planetary mass is assumed to be one Earth mass. The corresponding range of habitable zones is given by the following equation \citep{Kopparapu2014}:
\begin{equation}
	D_{HZ}=\left(\frac{L/L_{\odot}}{S_{eff}} \right)^{0.5} \ \mathrm{ AU},
\end{equation}
where $L/L_{\odot}$ is the luminosity of the star compared to the Sun.

We take the centre of the habitable zone as the orbital semi-major axis of the planet and calculate the strength of the signal\citep{Malbet2012}:
\begin{equation}
	\alpha=0.3\left(\frac{M_{P}}{M_{\oplus }}\right) \left(\frac{a}{1 \rm ~AU}\right) \left(\frac{M_{\star}}{M_{\odot}} \right)^{-1} \left(\frac{D}{10 \rm ~pc} \right)^{-1} \rm \mu as,
\end{equation}
where the $M_{P}$ and $M_{\star}$ are the mass of the planet and the host star, $a$ is the semi-major axis base on the habitable zone and $D$ is the distance of the host star. We take the classical signal-to-noise ratio SNR = 6 \citep{Brown2009} to estimate the upper limit of the allowable error due to photonic noise $\sigma =A/\rm SNR$. This will ultimately be used to determine the precision and time needed to observe Earth-like planets.

In the observation design, to ensure precision in the measurements, the CHES satellite will conduct a 2.5-hour calibration after each scientific observation of the target star, comprising 2 hours for optical aberration correction and 0.5 hours for focal plane calibration \citep{Ji2022}. Consequently, for a given observation target, optimizing the observation time for a single measurement and reducing the overall number of observations can minimize the calibration time share in the entire observation cycle, thereby enhancing observation efficiency. It is essential to strike a balance, ensuring an adequate number of observations to obtain a satisfactory set of stellar astrometric measurements.

\begin{figure}
	\centering
	\includegraphics[width=0.5\textwidth, angle=0]{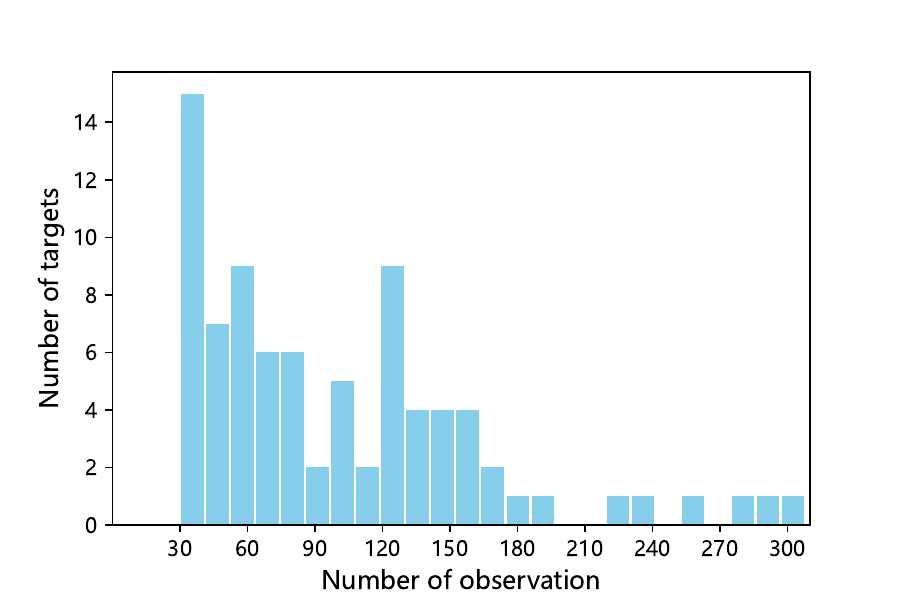}
	\caption{
		The distribution of the proposed number of observations for each target star is determined by various parameters. The average number of observations was approximately 100. The minimum required number of observations is set at 30, while the maximum number of observations reaches 308.}
	\label{number observation}
\end{figure}

The ultimate design entails observing each target star a minimum of 30 times during the 5-year observation period, with each scientific observation lasting no less than 0.5 hours. The distribution of the number of observations is presented in Fig~\ref{number observation}. The specific number of observations for each target star is detailed in the Appendix.

\section{Prioritization of target stars observation}
\label{sect:Prior}

We further filtered out the target stars that require less than about 300 observations based on the number of observations of the target stars in the previous section, totaling 94 target stars. Among them, there are 11 groups of two target stars in the same FOV, which will be observed at the same time, resulting in a total of 83 observation targets. For these targets, an observing strategy will be developed based on that one in the FOV that requires a higher number of observations.  Given that the time required for a single observation is no less than 3 hours, prioritization of target stars is essential to efficiently observe scientifically interesting ones within the designated operational timeframe. The prioritization of observations is determined by considering the following five factors:

\begin{enumerate}[label={\Roman*.}]

\item {D}: Distance. We intend to observe nearly 100 stars within 10 pc, and as a result, we will prioritize observations from the closest to the farthest in terms of distance. The value for distance prioritization ranges from 1.0 for the closest target star (* alf Cen A, 1.3459 pc \citep{Voges2000}) to 0.8 for the farthest target star (* i Boo A, 12.9450 pc \citep{GaiaCollaboration2018}). The other target stars will be assigned values between 0.8 and 1.0 based on their respective distances.

\item {M}: Magnitude. Considering the impact of photon noise, we analyze the effect of magnitude on observing priority, taking into account the illumination of the target star.
\begin{equation}
	\frac{E_2}{E_1}=10^{0.4\times\Delta M}
\end{equation}

The calculation of the effect of magnitude on observations is based on the illumination of a target star with a magnitude of 3. For targets with a magnitude less than or equal to 3, this factor is assigned a value of 1. For the target with the minimum magnitude (HD 32450, magnitude 9.6368 \citep{GaiaCollaboration2020}), this factor is assigned a value of 0.8. For other targets, the factor is assigned values in the range of 0.8 to 1.0, depending on the illumination.

\item {N}: Number of the reference stars. The number of reference stars affects both measurement error and the impact of photon noise. Reference stars in the FOV are generally selected with magnitudes not exceeding 13. For some target stars in a dense star field, there may be numerous reference stars. In actual observation, the brightest reference stars are chosen to measure the angular distance between them and the target star. In this prioritization, we consider cases with no more than 12 reference stars. For cases with more than 12, their influence on priority is equated to that of 12 reference stars. The values representing their influence on priority are as follows:

Value : $>12$:1.0; 8-11 : 0.9 ; 6-7 : 0.8 : $<6$ : 0.7

\item {E}: {Exoplanet. Kepler/K2 and TESS or ground-based telescopes, have detected a large number of exoplanets and candidates using transiting or radial velocity methods\footnote{https://exoplanetarchive.ipac.caltech.edu/docs/counts\_detail.html\label{fn2}}. The target stars selected for CHES include those that have been previously confirmed to have exoplanets or are tentative candidates. For instance, tau Cet has four identified planets \citep{Feng2017b}: tau Cet e, tau Cet f, tau Cet g, and tau Cet h. Other target stars with confirmed exoplanets and tentative candidates are listed in Table~\ref{Exoplanet}. The planetary systems in Table~\ref{Exoplanet} were all detected using the radial velocity method. Among them, HD 219134 b and HD 219134 c were also detected using the transit method. The CHES will further confirm the tentative candidates but also provide true planetary mass and three-dimensional orbits for the confirmed planets, thereby leading to a full understanding of planetary formation and evolution. The impact of the presence of exoplanets on priority is determined as follows:}

Value : Confirmed planets or tentative candidate: 1.0; Uncomfirmed: 0.8

\item {B}: Binary or more stars. This factor takes into account the possibility of using the same reference star in the same FOV for simultaneous exoplanet detection of the binary star, thereby improving observation efficiency. Additionally, observations of binary stars can yield high-precision astrometric data. The special case mentioned in Section~\ref{sect:mission} involving stars V* AK Lep and * gam Lep is also considered here as a binary star assignment to this factor.

Value : Binary or more stars: 1.0; single stars: 0.8

\end{enumerate}

The priority of each target star is determined by a combination of the five factors mentioned above, represented as ${rank index = D * M * N * E * B}$. The final priority is obtained by normalizing this value and is presented in the Appendix. A higher value indicates a higher priority for observation. Given the current number of established observations and the exposure time for a single observation, all observation targets can receive sufficient observation time. This prioritization of observation targets serves as a strategy for discarding certain observation targets in scenarios where additional observation tasks are introduced and there is insufficient observation time.

\begin{deluxetable*}{cccc}
	\tablenum{1}
	\tablecaption{Confirmed planets and tentative planets in CHES targets.\label{Exoplanet}}
	\tablehead{
		\colhead{Host name} & \colhead{Other name}&\colhead{Confirmed Planet} & \colhead{Tentative Planet}
	}
	\startdata
	61 Vir &- &61 Vir b (1,2); 61 Vir c (1,2) & 61 Vir d (2) \\
	HD 20794&e Eri & HD 20794 b (3,4); HD 20794 c (3,4); HD 20794 d (3,4) & HD 20794 e (4) \\
	tau Cet & -&tau Cet e (5); tau Cet f (5); tau Cet g (5); tau Cet h (5) & \begin{tabular}[c]{c}tau Cet b (6); tau Cet c (6);\\
		 tau Cet d (6) \end{tabular}\\
	HD 102365 &-& HD 102365 b (7) & - \\
	HD 147513 &-& HD 147513 b (8) & - \\
	HD 69830 &-& HD 69830 b (2,9); HD 69830 c (2,9); HD 69830 d (2,9) & - \\
	eps Eri &-& eps Eri b (2,10,11,12,13) & - \\
	eps Ind &-& eps Ind A b (14,15) & - \\
	HD 26965& omi02 Eri & - & HD 26965 b (16) \\
	HD 131977 &-& - & HD 131977 b (17) \\
	HD 219134 &-& \begin{tabular}[c]{c}HD 219134 b (2,18,19,20,21,22); HD 219134 c (2,18,19,21,22); \\
		HD 219134 d (2,18,19,22); HD 219134 f (2,18,19,22);\\
		HD 219134 h (2,22,23)\end{tabular} & HD 219134 g (22) \\
	HD 192310 &-& HD 192310 b (2,4,24) & HD 192310 c (4) \\
	GJ 86& HD 13445 & GJ 86 b (25,26,27) & - \\
	HD 3651& 54 Psc & HD 3651 b (2,27,28,29,30) & - \\
	HD 85512 &-& HD 85512 b (4,25) & - \\
	HD 40307 &-& \begin{tabular}[c]{c}HD 40307 b (31,32,33); HD 40307 c (31,32,33); \\HD 40307 d (31,32,33);
		 HD 40307 f (31,32,33);\\ HD 40307 g (31,32)\end{tabular} & - \\
	lam Ser &-& - & lam Ser b (2) \\
	HD 115404&LHS 2713 &-  & \begin{tabular}[c]{c}HD 115404 A b (34);\\ HD 115404 A c (34) \end{tabular}\\
	\enddata
	\tablecomments{References:(1)\cite{Vogt2010}; (2)\cite{Rosenthal2021}; (3)\cite{Feng2017a}; (4)\cite{Pepe2011}; (5)\cite{Feng2017b}; (6)\cite{Dumusqu2017}; (7)\cite{Tinney2011}; (8)\cite{Mayor2004}; (9)\cite{Lovis2006}; (10)\cite{Llop-Sayson2021}; (11)\cite{Mawet2019}; (12)\cite{Benedict2006}; (13)\cite{Hatzes2000}; (14)\cite{Feng2019}; (15)\cite{Feng2023}; (16)\cite{Ma2018}; (17)\cite{GaiaCollaboration2020}; (18)\cite{Gillon2017}; (19)\cite{Motalebi2015}; (20)\cite{Kokori2023}; (21)\cite{Seager2021}; (22)\cite{Vogt2015}; (23)\cite{Johnson2016}; (24)\cite{Howard2011}; (25)\cite{Stassun2017}; (26)\cite{Butler2001}; (27)\cite{Butler2006}; (28)\cite{Wittenmyer2019}; (29)\cite{Wittenmyer2007}; (30)\cite{Wittenmyer2009}; (31)\cite{Tuomi2013}; (32)\cite{Brasser2014}; (33)\cite{Daz2016}; (34)\cite{Feng2022}}
\end{deluxetable*}

\section{Strategy for observation}
\label{sect:Strategy}
Sky survey telescopes typically follow an established scanning law to observe targets entering their FOV \citep{GaiaCollaboration2016}. Given that the observation will encompass the entire celestial sphere, the allocated observation time and the number of observations for each target are limited. Furthermore, adhering to the principles of the scanning law, the time intervals between consecutive observations are relatively extended. This temporal arrangement may result in the oversight of certain periodic signals. Specifically, signals associated with the gravitational effects of exoplanets on the target star may be absent from the observations.

In contrast, CHES focuses on pre-screened candidate targets. Its observation accuracy is higher, and the duration of a single observation is longer. Therefore, we have tailored an observation strategy specifically for CHES. This strategy aims to ensure that the number of observations for each target star is sufficient, while maintaining a uniform time interval between observations. This approach guarantees that the observations capture accurate information about position changes, allowing for the deduction of information about surrounding planets, as well as the masses and orbits of habitable planets.

As detailed in Section ~\ref{sect:mission}, the CHES satellite conducts observations in two modes: conventional mode and revisited mode. The two types of conventional modes are interspersed over the five-year observation period and constitute the majority of the observing time. The first type is utilized during the first, third, and fifth years of observations, while the second type is employed for the remaining two years. In both types of conventional modes, the range of observations is fixed relative to the position of the Sun. Consequently, as the satellite trails the Earth in its orbit around L2, each region will be observed twice during a year of observation.

\begin{figure}
	\centering
	\includegraphics[width=0.4\textwidth, angle=0]{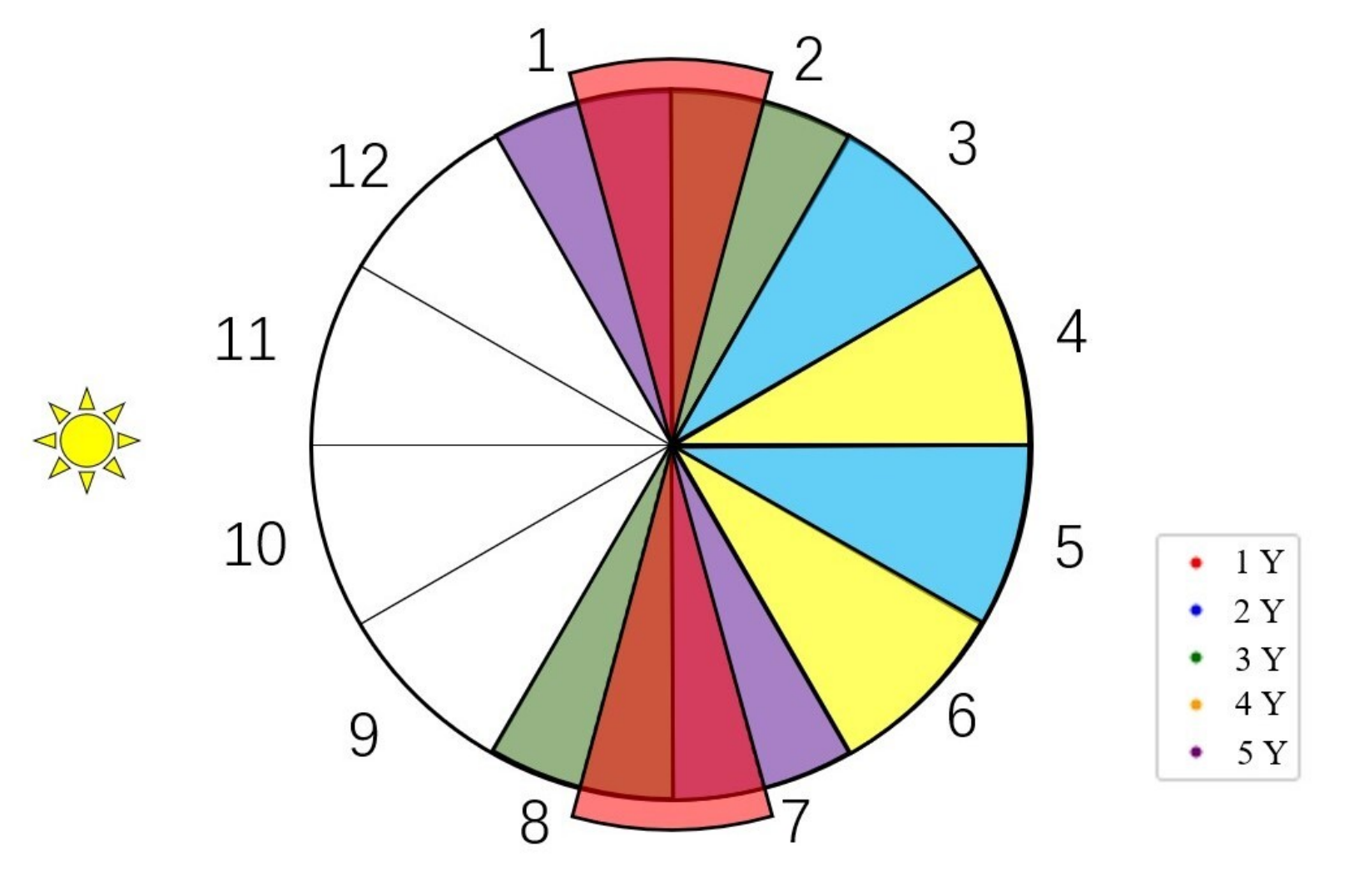}
	\caption{Observed ranges under different years at each moment. In the 1st, 3rd, and 5th years, conventional mode type one will be used. In the 2nd, 4th years, conventional mode type two will be used. To avoid the effects of sunlight, no observations will be conducted in regions 9-12.}
	\label{ob1}
\end{figure}

Figure~\ref{ob1} illustrates the regions to be observed at each time during every year. The figure provides a top view of the ecliptic plane, with the Sun's ecliptic determining the range of ecliptic longitudes observable by satellites. To avoid direct sunlight, the detector observation direction should be at least $60^{\circ}$ away from the direction of the sun, denoted as regions 9-12 in the figure. In the first type of conventional mode, the telescope rotates along the great circle passing through the north and south ecliptic poles, with the observation regions represented in red, green, and purple. In the second type of conventional mode, the telescope observes along two semicircles perpendicular to the ecliptic plane, connected at the north and south poles of the ecliptic, forming an angle of 60 degrees between them. The observation regions are depicted in blue and yellow.

For the annual parallax of a target star, the maximum parallax occurs when the difference between its ecliptic longitude and that of the Sun in the station-centered ecliptic coordinate system is $90^{\circ}$. In the first year, the red region in Fig~\ref{ob1} is selected and overlapped with the purple and green regions, aiming to achieve more observations at the maximum parallax, thereby enhancing the accuracy of the parallax measurement.

\begin{figure*}
	\centering
	\includegraphics[width=0.8\textwidth, angle=0]{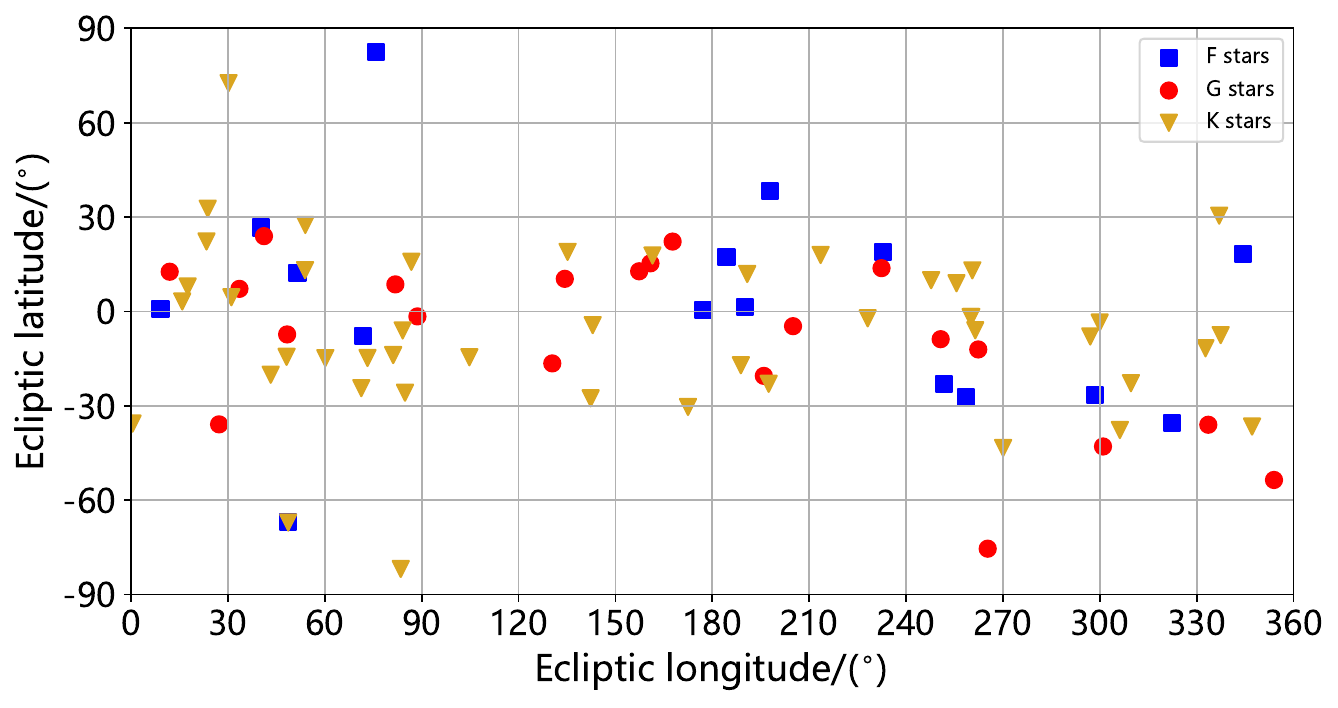}
	\caption{Distribution of targets in ecliptic coordinates. The distribution is not homogeneous, leading to the need to use the re-observation mode to make adjustments in the observation time.}
	\label{ecliptic coordinates distribution}
\end{figure*}

In the course of the observation, the satellite is designed to operate with an observing range of 30 degrees in ecliptic longitude. As an example of the regions of observations in the first year, when the ecliptic longitude of the Sun in the station-centered ecliptic coordinate system is $0^{\circ}$, the range of observation can be $75^{\circ} - 105^{\circ}$ and $255^{\circ} - 285^{\circ}$ in the first year. The satellite is in the vicinity of the L2 point, and the plane of observation is perpendicular to the ecliptic plane, so stars with close ecliptic longitudes in ecliptic coordinates will be observed in close proximity. Figure~\ref{ecliptic coordinates distribution} shows the distribution of targets in ecliptic coordinates.

\begin{figure*}
	\centering
	\includegraphics[width=0.8\textwidth, angle=0]{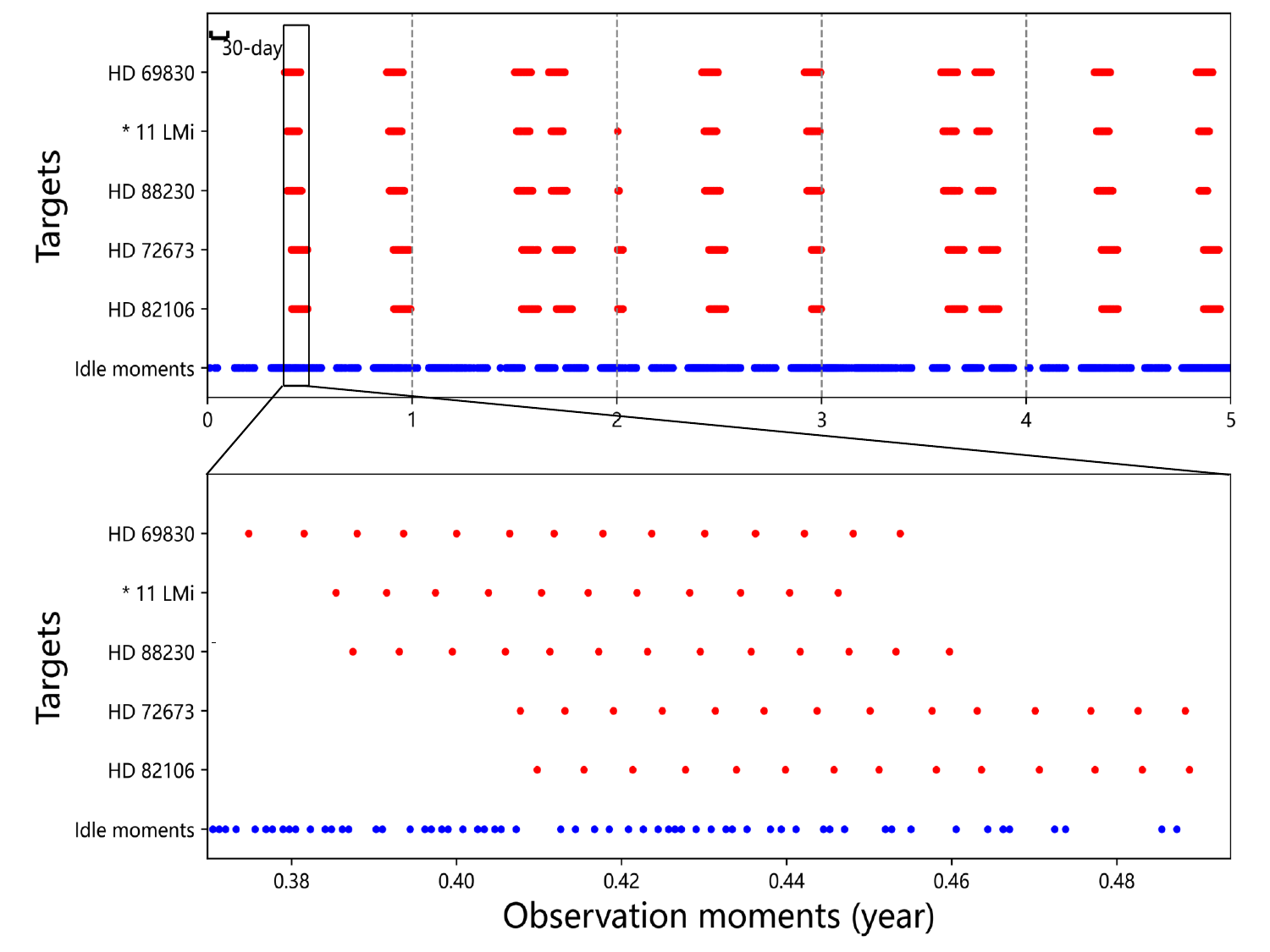}
	\caption{
		Distribution of observation moments for targets and idle. The five target stars are distributed around 135 degrees of ecliptic longitude. The figure below is a partial enlargement of the black box in the above figure. The order of observation is determined by the position of the target star and the number of observations. At this ecliptic longitude, the distribution of target stars is also relatively sparse, providing idle time in between observations. This idle time can be used to intersperse with other observing tasks.}
	\label{fig:distribution of moment}
\end{figure*}
The satellite moves with the rotation of the Earth at the L2 point, and the range of its observation will also change with time. This allows the regions of observation to cover the entire celestial sphere within the first six months. Since the observations are within a certain range, any target star can be observed for at least two time periods per year, totaling about two months. Within each time period, to ensure that the observation time is as uniform as possible, each observation for a single target star is separated by a certain amount of time. We adopt the following scheme to determine the observation interval for each target star:

\begin{eqnarray}
     \label{N}
        N_{per}=\frac{\lambda_{width}}{v_{rot}} N_{t},\\
		N_{int}=\frac{N_{per}}{N_{obs}/10},
\end{eqnarray}
where ${\lambda_{width}}$ represents the observing range of the telescope in ecliptic longitude, set to $30^{\circ}$ in this paper. $v_{rot}$ stands for the speed of the telescope as it rotates along the ecliptic with the Earth at point L2. $N_{t}$ denotes the average number of observations per day. $N_{per}$, $N_{obs}$, and $N_{int}$ represent the total number of observations in the observing range, the total number of observations per star, and the number of intervals between observations, respectively. In a 5-year observation period, each star has 10 subperiods, so the number of observations required in each subperiod is $N_{obs}/10$.

Ultimately, the number and duration of observations of each target star are distributed as uniformly as possible over the 10 observation periods within the 5-year operating time. Meanwhile, to ensure the efficiency of the observation, the positions of the target stars on the celestial sphere during two adjacent observations need to be as close as possible, so as to avoid the detector's pointing to make reciprocal rotations over a large area.

In addition to the conventional mode, there is the previously mentioned revisited mode. The situations that require the use of the revisited mode are twofold.

\begin{enumerate}[label={\Roman*.}]
	\item  According to the Section~\ref{sect:Photon}, when some target stars require a large number of observations, a sufficient number of observations cannot be accomplished within the time period in which they can be observed, while ensuring that each observation is separated by a certain period of time.

	\item  As shown in Fig~\ref{ecliptic coordinates distribution}, the distribution of target stars at some ecliptic longitudes (e.g., $45^{\circ}$ and $255^{\circ}$) is relatively dense. In this case, although the number of observations required for each of these target stars may be small, the total number of observations required for the target stars within this range is too large. Since the detector has limited time to cover this range, the number of observations is limited, which may result in some target stars not being fully observed for a sufficient number of times during the corresponding observing period.
\end{enumerate}

In both cases, the revisited mode is employed to conduct additional observations during idle moments of the telescope. As mentioned earlier, there are ecliptic longitudes where the distribution of target stars is dense, and conversely, there are some ecliptic longitudes where the distribution of target stars is sparser (e.g., $115^{\circ}$). When observing these target stars, there are telescope idle moments since each star has its own observation interval. Multiple observations of the same target star with a long orbital period in a short period are meaningless. During this idle moments, the telescope can be used for checking and filling in the gaps of the target stars observation due to above two points, but also for other scientific programs.

Based on the single observation time plan described in Section~\ref{sect:Photon} and the observation program described in this section, we formulate the final observation strategy for the target stars. Examples of some target stars with close ecliptic longitudes are shown in Fig~\ref{fig:distribution of moment}. The idle moments of the telescope are shown at the bottom of Fig~\ref{fig:distribution of moment}. In this observation strategy, the total duration of the satellite's 5-year operating time is 29,220 hours ($16~h \times 365.25 \times 5$), with approximately 25,120 hours dedicated to the total observation time on the target stars (about 86 percent of the entire time), as shown in Table~\ref{time}. The duration available for other observation missions is about 4,100 hours. These times can be distributed at any point during the 5-year operating time by further adjustments to the observing strategy to ensure that the basic observation task is satisfactorily completed, along with other observing tasks.

\section{Impact of observing strategies on scientific missions}
\label{sect:Impact}
For the observation strategy, the CHES observation time series is not uniform. The observation moments for targets vary from year to year, and the distribution of observation times shows segmentation. For targets that require more observations, the observations are more discrete. Nevertheless, in the 5-year observation, the observations of each year are complementary to each other, satisfying the needs of scientific observations.

As to the observation design, the target star is positioned at the center of the FOV, ensuring that the results of multiple observations reveal relative changes in the position of the reference star. The celestial motion of the star is a cumulative effect of various motions, and alterations are manifested in the relative proper motion and relative parallax between the target star and the reference star, along with the influence of other factors on the star's position.

\subsection{Proper motion and parallax}

\begin{figure*}
	\plotone{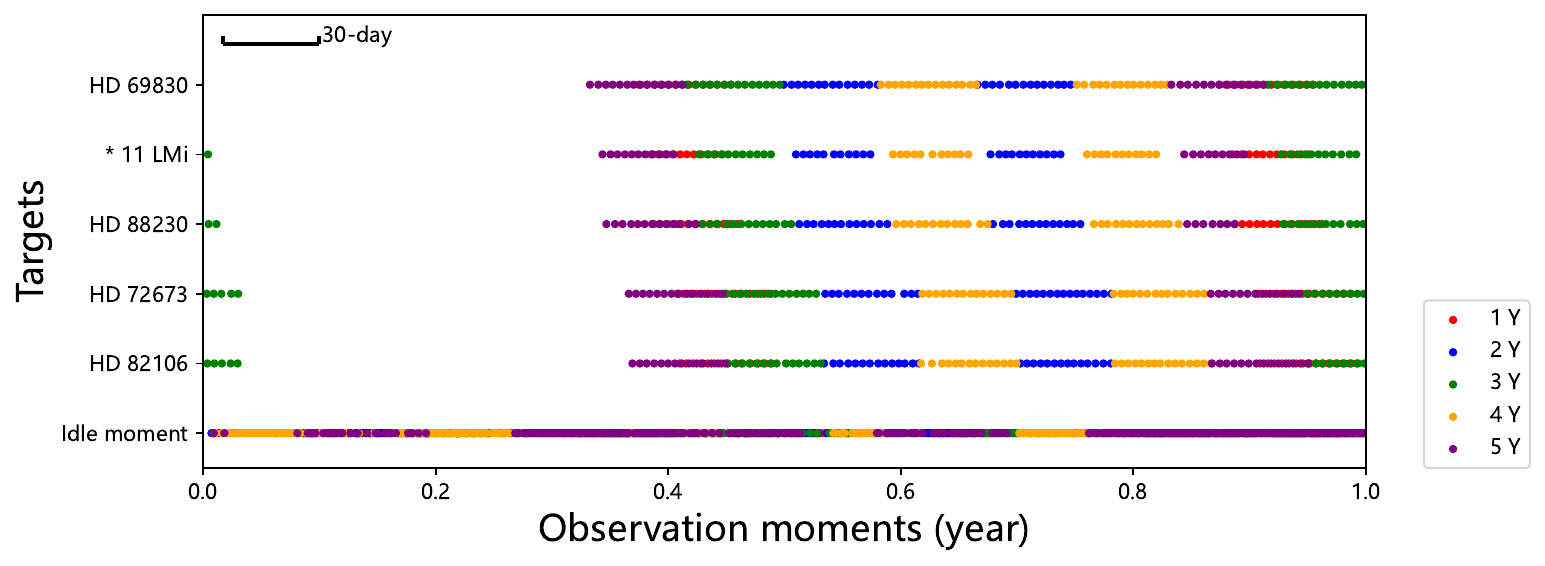}
	\caption{Distribution of observation moments for each target (five for illustration) overlapped in one year.}
	\label{fig:distribution of moment1yr}
\end{figure*}

\begin{figure*}
	\centering
	\includegraphics[width=0.7\textwidth, angle=0]{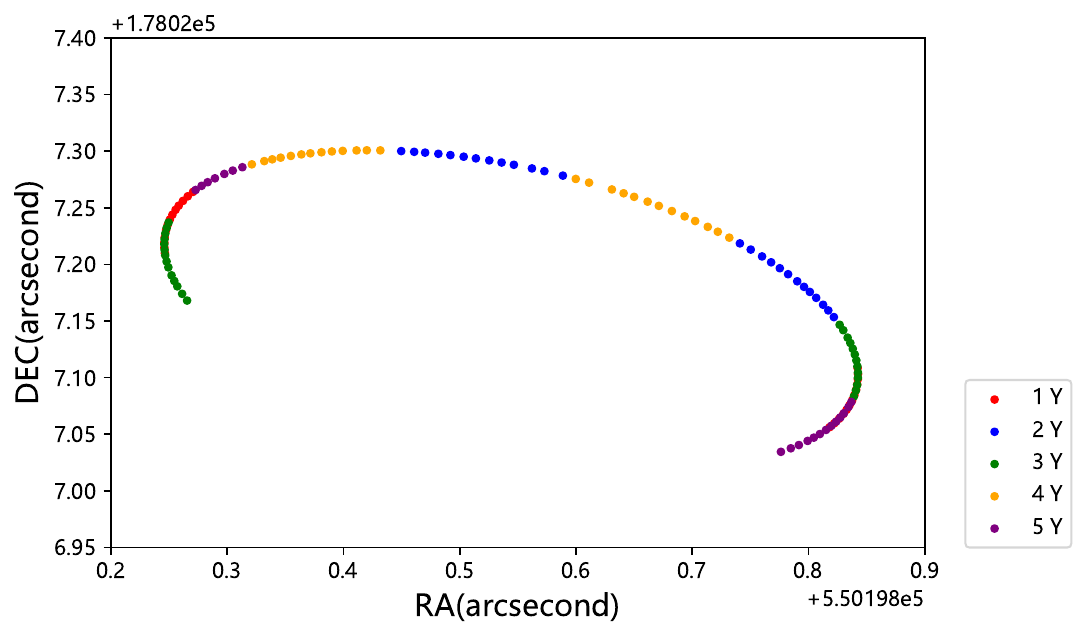}
	\caption{Observable part of the parallax ellipse (Target: HD 88230 \citep{GaiaCollaboration2020})}
	\label{fig:parallax}
\end{figure*}

We analyzed the distribution of observation strategies in Fig~\ref{fig:distribution of moment} by consolidating the overlap of observations for each year into a single year, and the results are illustrated in Fig~\ref{fig:distribution of moment1yr}. Various colors in the figure denote observations in different years. Maintaining an angle of at least 60 degrees between the observation direction and the Sun is essential to avoid the Sun's influence. Consequently, for any target star, there must be a one-third of the year during which it cannot be observed, represented by the blank time period in Fig~\ref{fig:distribution of moment1yr}.

While the blank time period implies that only two-thirds of the parallax ellipse of the target star will be observed, it is sufficient to fit the parallax with a high degree of precision. According to the observing strategy outlined in the previous section, the observable parallax ellipse can be entirely captured during a 5-year period, with additional observations concentrated when the target star is positioned near its parallax maximum (i.e., the region where the first, third, and fifth years of observations overlap). The observed part is depicted in Fig~\ref{fig:parallax}. The proper motion has been eliminated, retaining only the parallax of the target star.

\subsection{Coverage of planetary orbit period}
The mission's primary objective is to acquire data on the orbits of planets exhibiting periodic motion around their stars. Consequently, the mission's efficacy in detecting these orbits will be crucial. The capability to observe the periodic motion of the planets will directly impact the precision of orbit fitting. In the observing strategy outlined in the previous section, observations of any star are not continuous. Thus, it is essential to verify the mission's capacity to cover the orbital period of the planets in our observations.

\begin{figure}[ht]
	\plotone{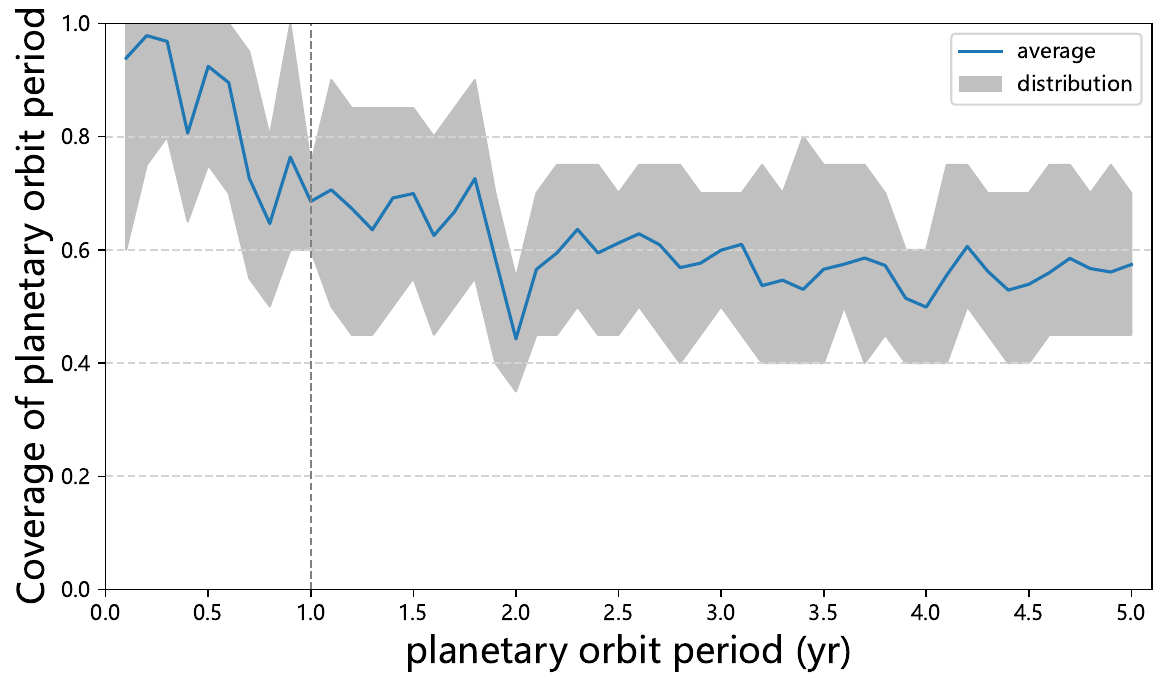}
	\caption{Coverage of planetary orbit period. The blue line indicates the mean value, while the grey ranges indicate the distribution of period coverage under each planetary period for all target stars.}
	\label{fig:Coverage}
\end{figure}

We divided the orbital period into 20 parts, each representing 0.05 of the period. A period is considered observed if the observation time for the target star falls within a specific part. Using the observation moments for each target star from the aforementioned observing strategy, we simulated all target stars for planets with varying periods (ranging from 0.1 to 5 years). The distribution of the final coverage is illustrated in Fig~\ref{fig:Coverage}, with the blue line representing the average value. The coverage of planetary orbital periods consistently exceeds 50 percent and shows an increase with the number of observations, although the changes are not substantial. Our primary focus is on detecting Earth-like planets, i.e., those with periods around 1 year. The coverage of their periods ranges from approximately 60 to 75 percent.

\subsection{Simulation solution of planetary parameters}
\begin{figure*}
	\gridline{\fig{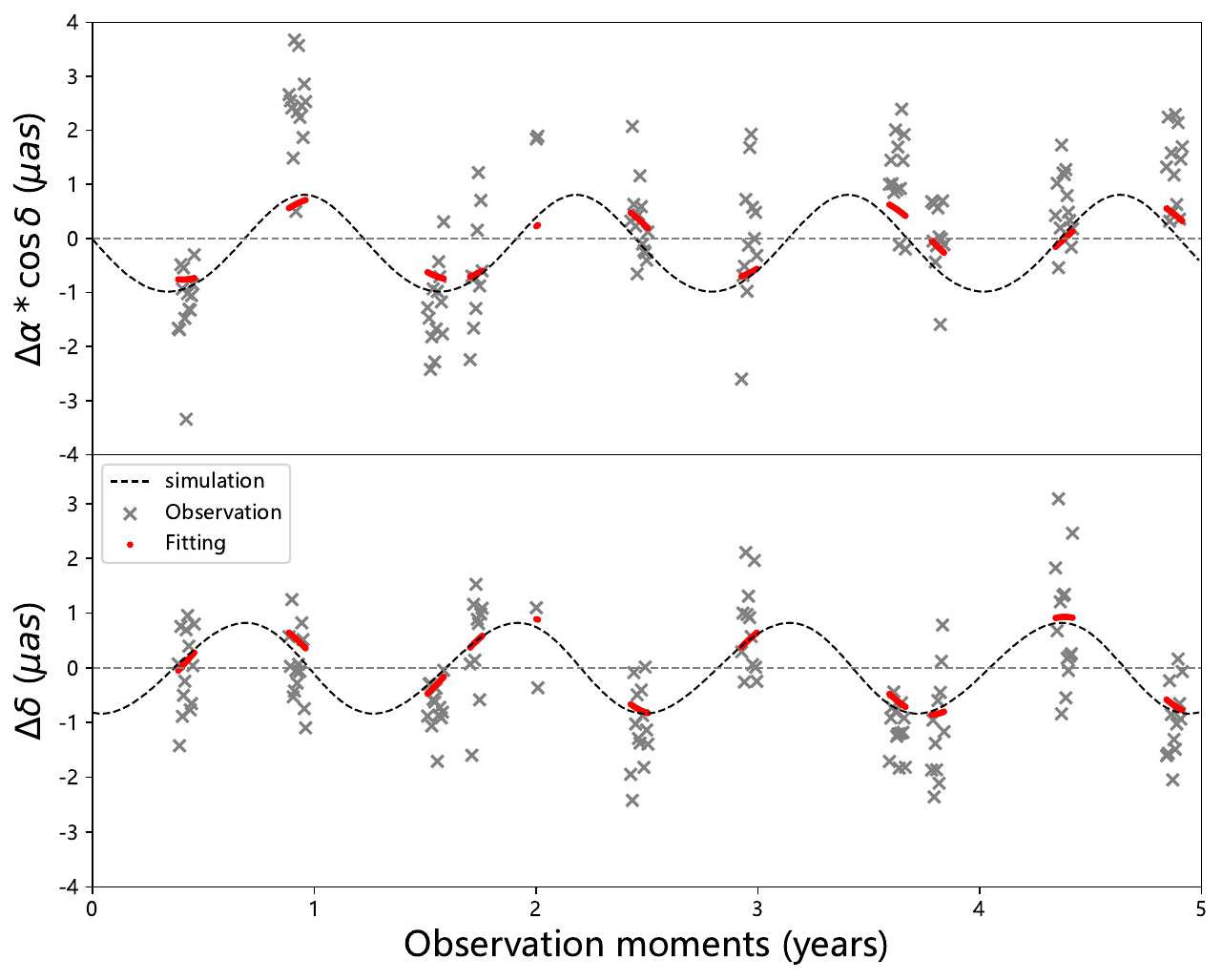}{0.5\textwidth}{(a)}
		\fig{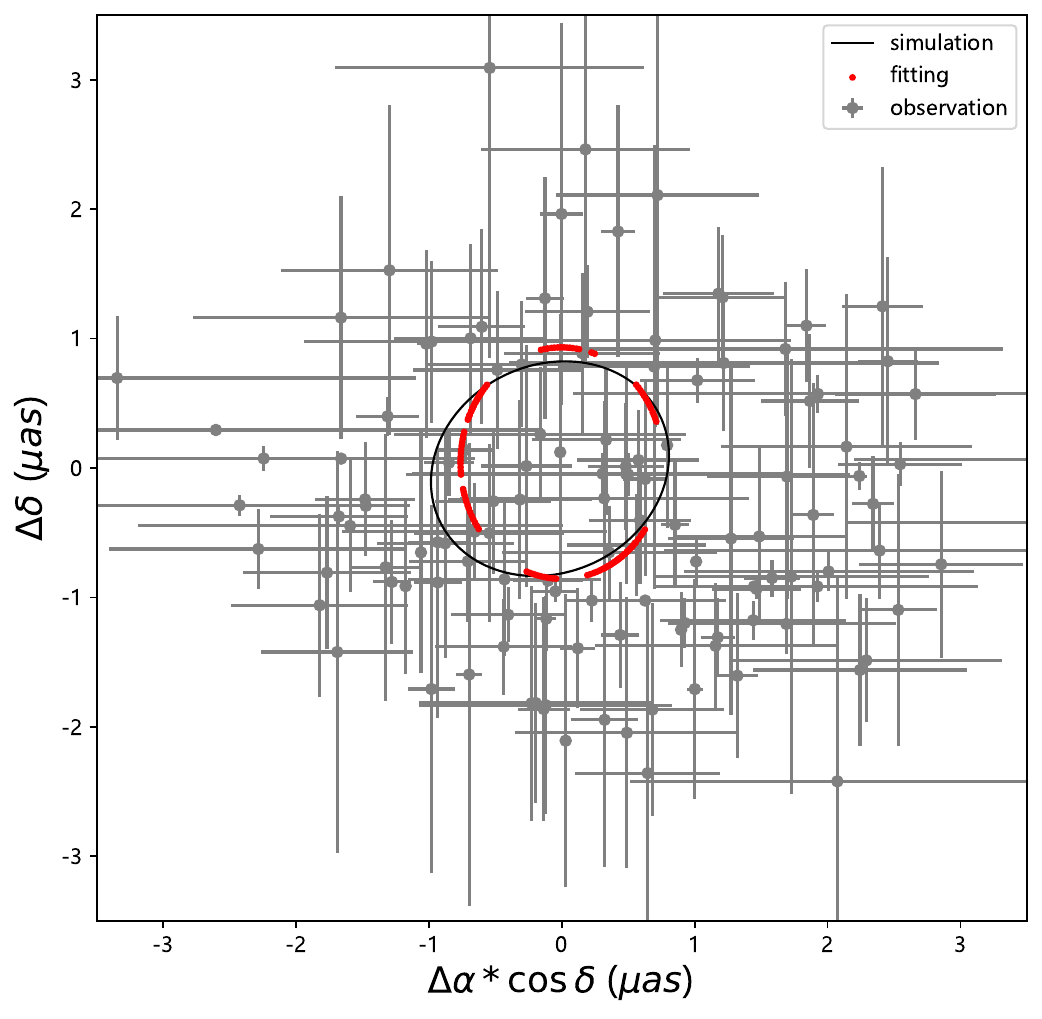}{0.4\textwidth}{(b)}
	}
	
	\caption{
		The black dotted line represent the simulated periodic motion of the target star caused by the presence of planet around it. The grey forks in (a) and dots with error bars in (b) depict the residual motion that remains after eliminating the relative proper motion and parallax of the target and reference stars in the fit based on the relative measurement method. The red dots illustrate the outcome of the fit to this periodic motion. \emph{Left panel}: Component at right ascension and declination. \emph{Right panel}: Periodic motion of the target star on the celestial sphere.}
	\label{fig:radec}
\end{figure*}

\begin{figure*}
	\gridline{\fig{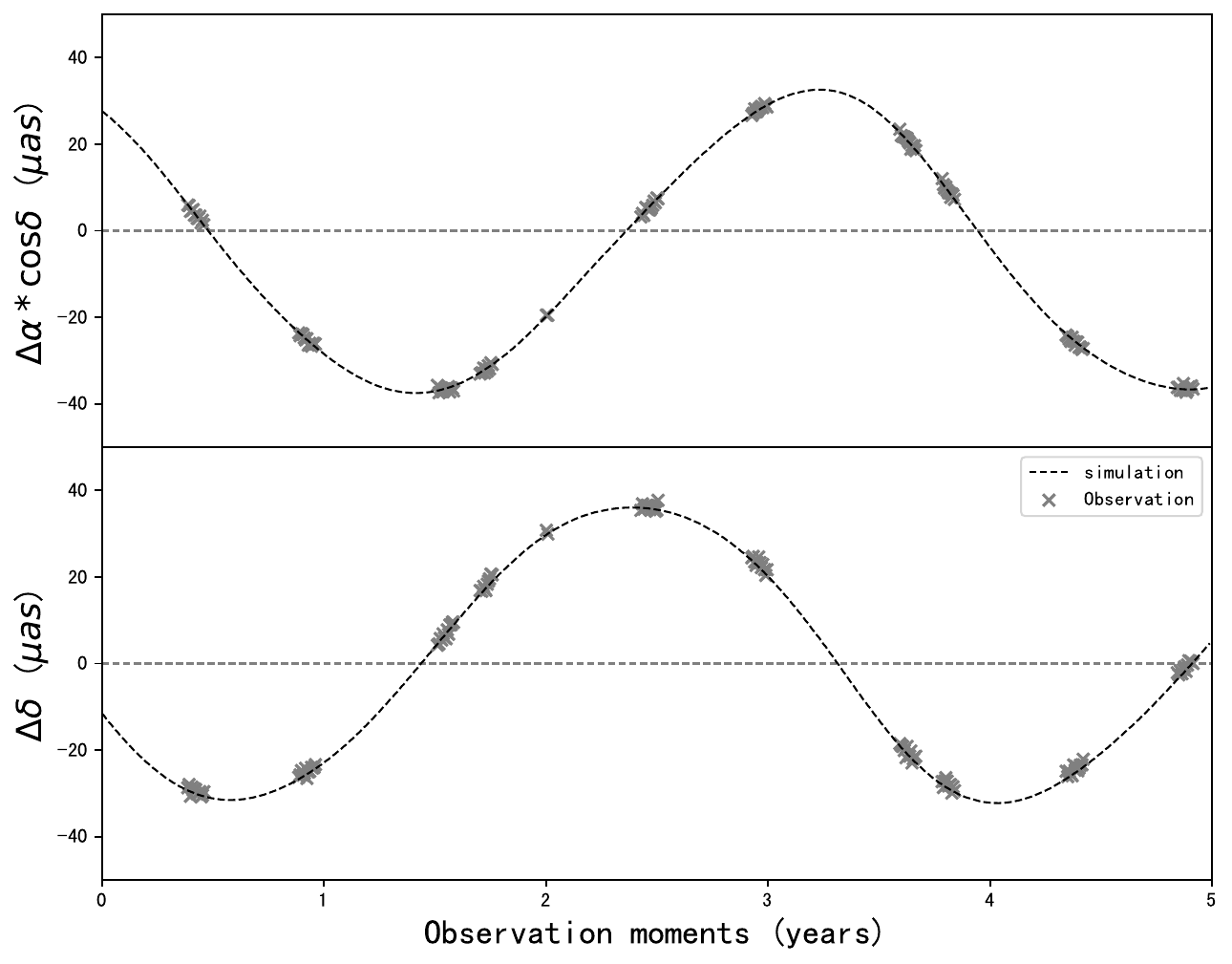}{0.5\textwidth}{(a)}
		\fig{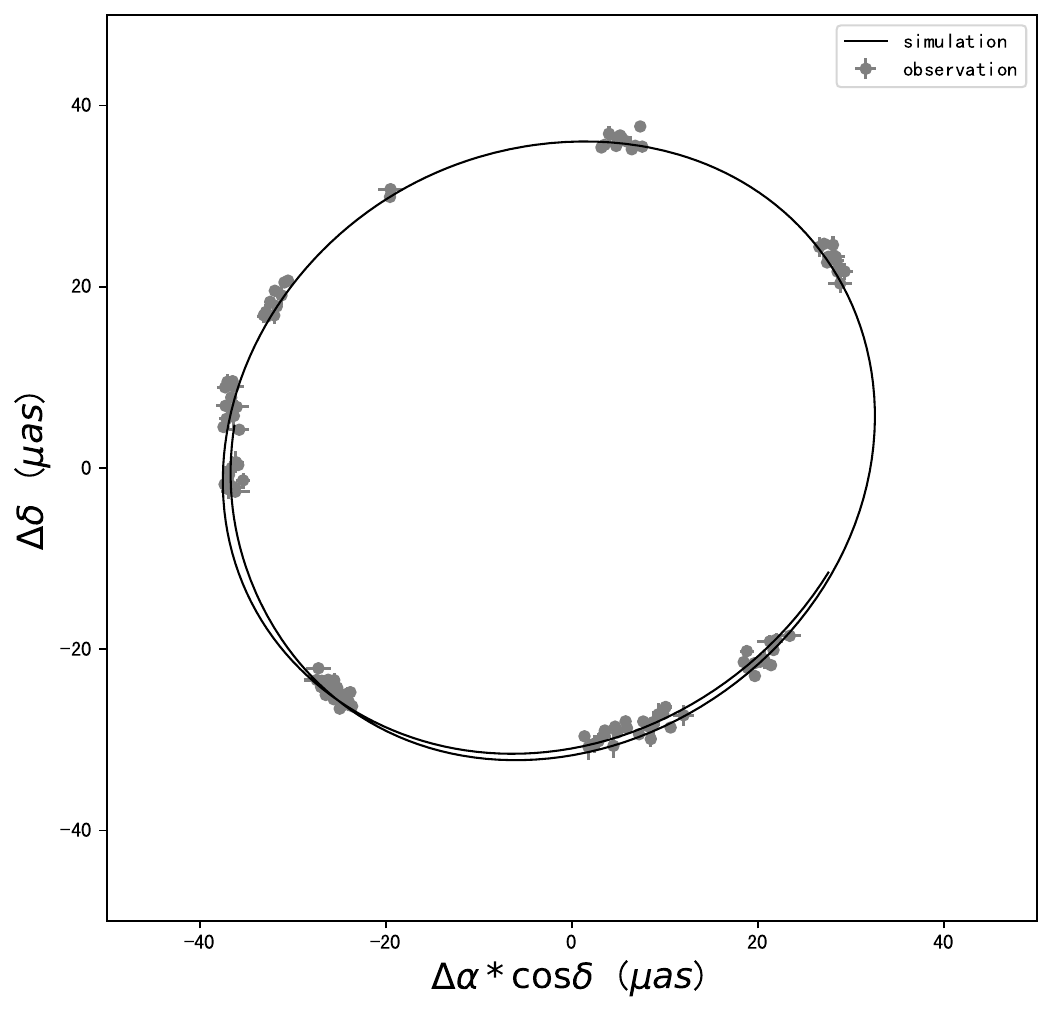}{0.42\textwidth}{(b)}
	}
	
	\caption{
		The black dotted line represent the simulated periodic motion of the target star caused by the presence of both planets around it. The grey forks in (a) and dots with error bars in (b) depict the residual motion that remains after eliminating the relative proper motion and parallax of the target and reference stars in the fit with the relative measurement method.  \emph{Left panel}: Component at right ascension and declination. \emph{Right panel}: Periodic motion of the target star on the celestial sphere.}
	\label{fig:radec_D}
\end{figure*}

\begin{figure*}
	\gridline{\fig{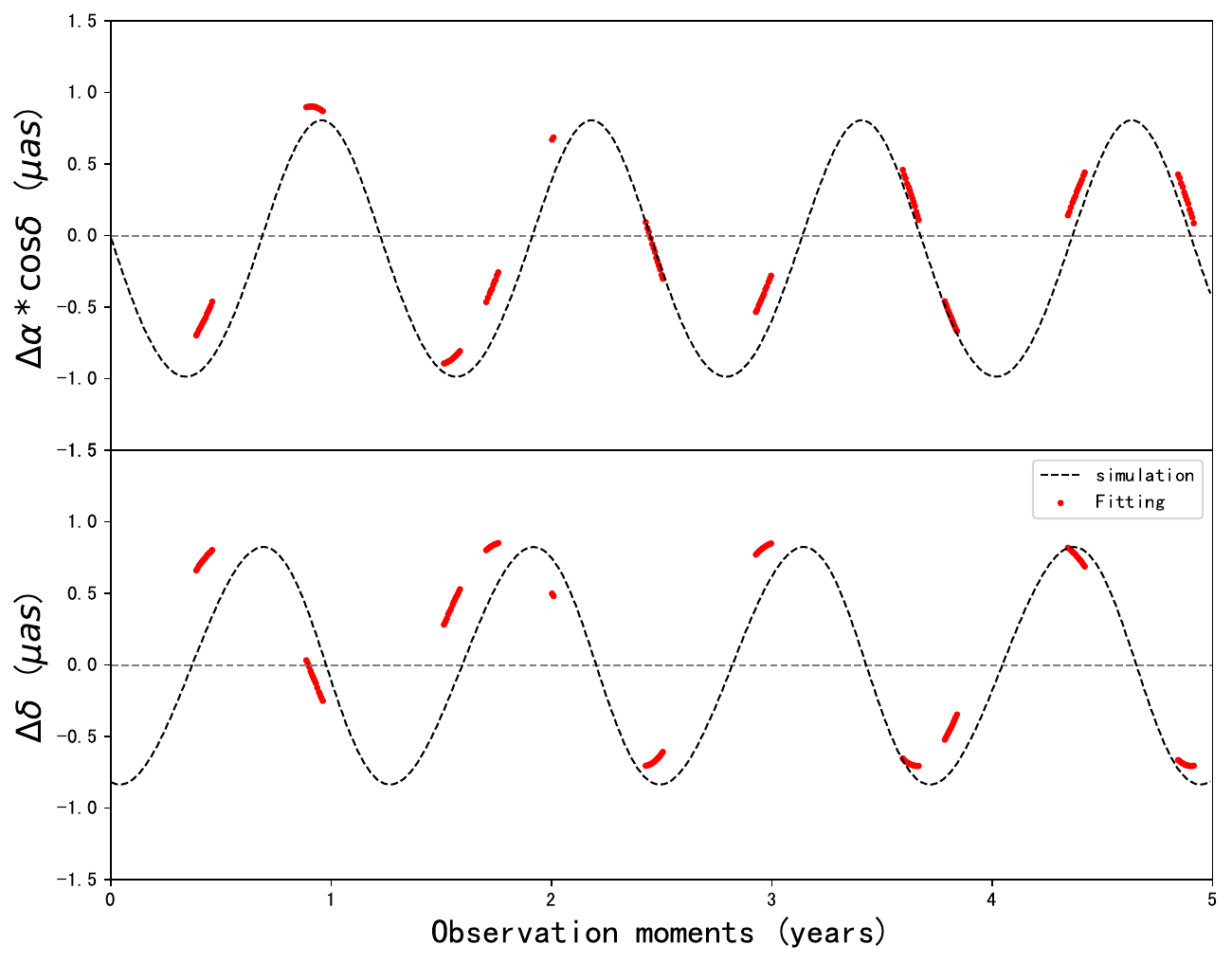}{0.5\textwidth}{(a)}
		\fig{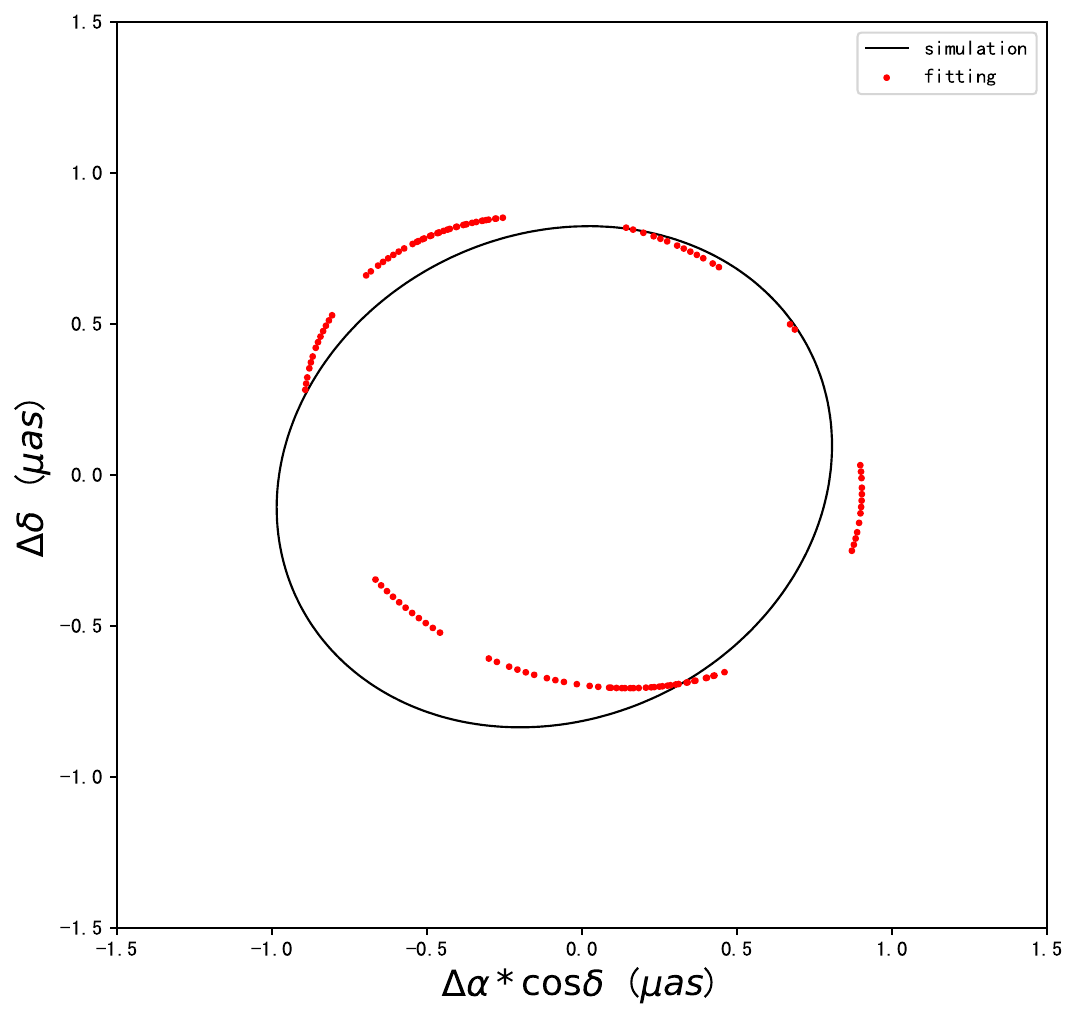}{0.42\textwidth}{(b)}
	}
	\gridline{\fig{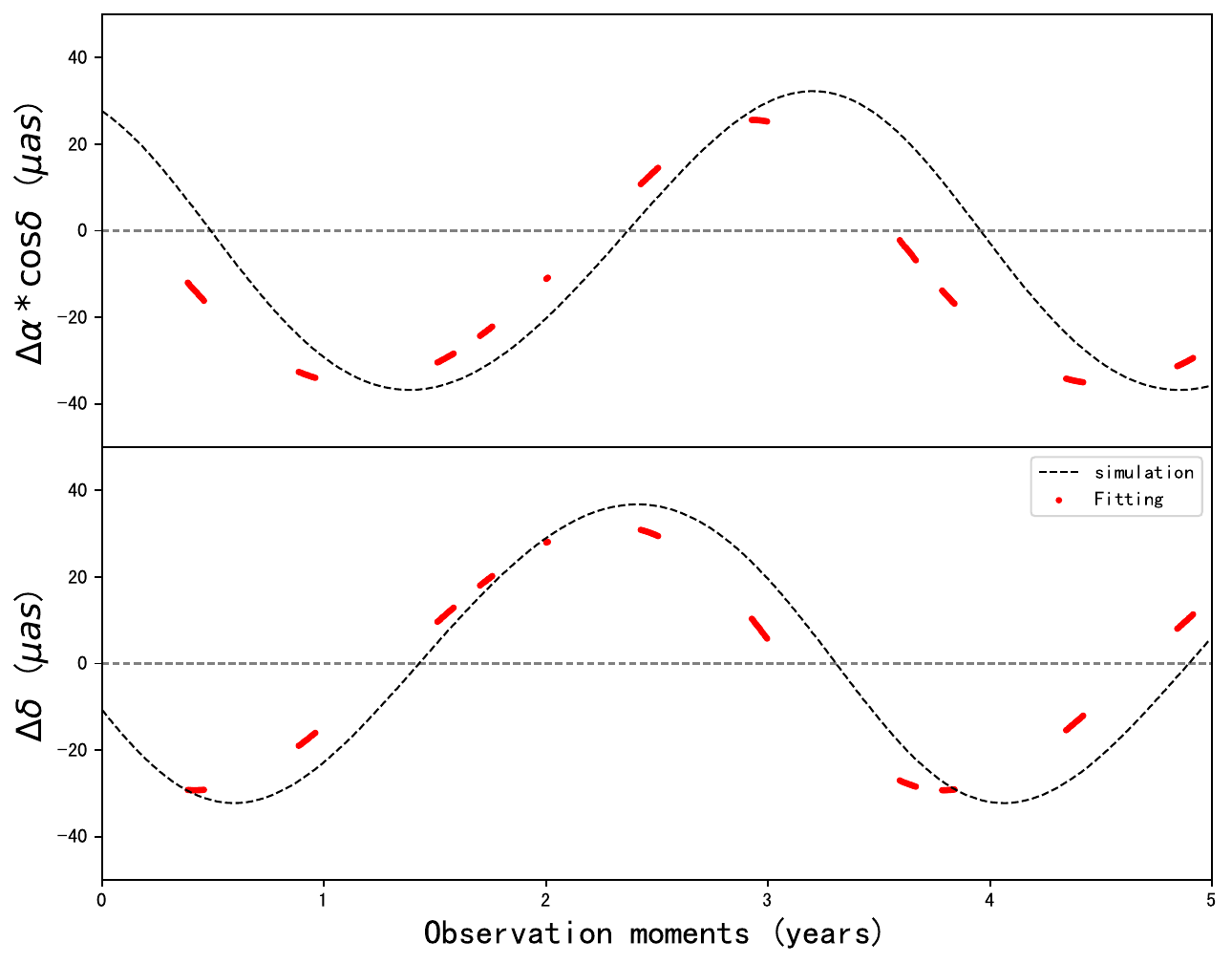}{0.5\textwidth}{(c)}
		\fig{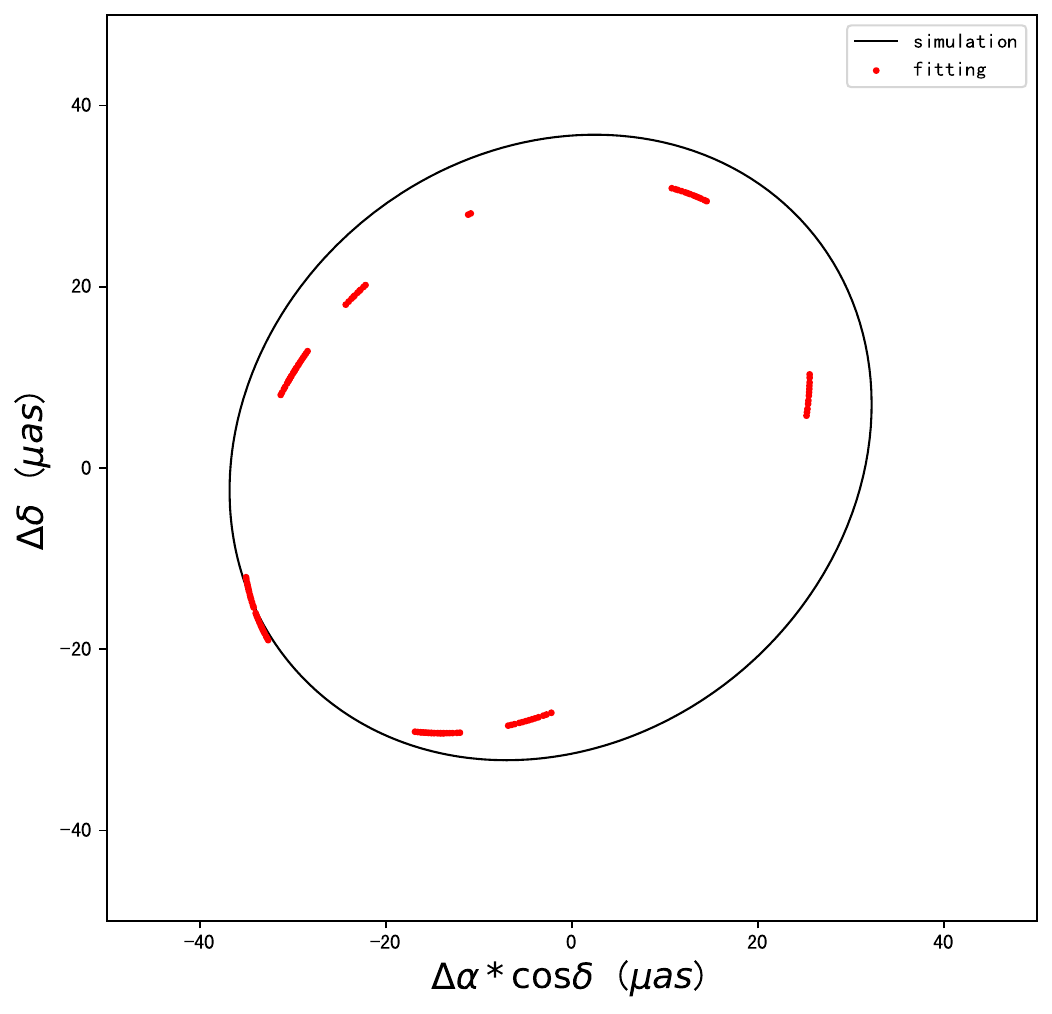}{0.42\textwidth}{(d)}
	}
	\caption{
		The black dotted line in the upper and lower panels represent the simulated periodic motion of the target star caused by the presence of two planets around it,  respectively. (a) and (b) correspond to the inner Earth-like planets. (c) and (d) correspond to the outer planet. The red dots  illustrate the outcome of the fit to these periodic motion. \emph{Left panel}: Component at right ascension and declination. \emph{Right panel}: Periodic motion of the target star on the celestial sphere.}
	\label{fig:radec_D2}
\end{figure*}
We chose HD 88230 \citep{GaiaCollaboration2020} as an illustrative example and conducted two simulations incorporating relevant parameters. The first example considers the scenario of an Earth-like planet at 1 AU from the star, while the second run is related to an additional planet with a mass 20 times that of Earth located at 2 AU. During these simulations, the target star undergoes periodic motion induced by the gravitational influence of the planets, resulting in positional variations at the micro-arcsecond level. Following the observing strategy outlined in this work, we simulated the FOV of the target star as observed by the detector from the L2 point at various time points. This field of view encompasses positional information of both the target and reference stars. Subsequently, we employed a fitting procedure on the angular distance data, incorporating a Gaussian error based on the detection capability.

As outlined in the preceding section, this mission relies on the principle of relative measurement to capture the periodic motion of the target star. Consequently, during the observation and data processing of the target star and the reference star, it is unnecessary to acquire precise position information of the target star. Simultaneously, obtaining position information for the target star with the required high accuracy (on the order of micro-arcseconds) is not feasible.

Based on the fitting method of relative measurements \citep{Bao2024B}, we fit the relative proper motions, and the parallaxes. The residual represents the periodic motion of the target star due to the presence of the planet. The orbital fit to this residual yields the orbital parameters. Figure~\ref{fig:radec} shows the results of the orbital fitting in the right ascension and declination directions, along with the simulated periodic motion of the star based on the observed moments from the observing strategy. Figure~\ref{fig:radec_D} and \ref{fig:radec_D2} illustrate the presence of two planets orbiting the star. Table~\ref{Fitting} summarizes the fitting parameters for one-planet and two-planet systems, respectively, along with the initial parameters, indicating that the fitted values are in close to the given data.

\begin{deluxetable*}{cccc}
	\tablenum{2}
	\tablecaption{Time allocation in CHES mission.\label{time}}
	\tablewidth{0pt}
	\tablehead{
		\colhead{Program} 	& Observation mode 	& \colhead{Used time (h)} &\colhead{Mission fraction}
	}
	\startdata
	CHES targets     & Conventional mode	&21451	&0.734   			\\
	& 	Revisited mode	&3669	&0.126\\
	&	Total	&25120	&0.860\\
	Open time&	& 4100	&0.140    		\\			
	Overall	&	&29220	&1.00	
	\enddata
\end{deluxetable*}

\section{Conclusions and Discussions}
In this work, we conduct an extensive study of the optimization of observation strategy for the target stars in the CHES mission. The primary objective of CHES is to identify and characterize potentially habitable Earth-like planets or super-Earths orbiting approximately 100 solar-type stars within a distance of 10 parsecs from our solar system. CHES accomplishes high-precision astrometric measurements by tracking angular distance variations between the target star and reference stars. A thorough analysis of relevant parameters for both target and reference stars has been carried out to determine required observation accuracy, the number of observations required, and priority assignment for each target star.

Here two observation modes of the observing strategy involve adopting different observation areas in different years, and measures such as peaks and troughs are adjusted to ensure a sufficient number of observations of all target stars within a 5-year period. With this observing strategy, every target star will be observed at least once within a six-month period. Over the course of the 5-year period, a total of 29,220 hours are available for scientific observations, with 25,120 hours (86 percent) designated for observing target stars, and the remaining 4,100 hours allocated to other programs. Each target star is assigned no fewer than 30 observations, with a maximum of 308 observations and an average of approximately 100 observations.  Reflecting the motion of a star by the variation of the angular distance between the reference star and the target star, CHES is not limited to the detection of terrestrial planets but can also detect factors affecting the reflex motion of the star, such as the motion of a binary star and its evolution, and the detection of black holes. The design of the observing strategy allows for the flexible allocation of time for other observing tasks, ensuring that CHES can achieve its utmost scientific goals while completing additional observation tasks.

There is scope for optimization within this observation strategy. During observations, to mitigate the impact of direct sunlight on the detector, the target star is discontinued from observation when the angle between the sun and the target star falls below a certain threshold. In the formulation of this strategy, this threshold angle is set to 60 degrees. Consequently, the target star remains unobserved for one third of the year, as shown in Figure~\ref{fig:parallax}. This angle is determined by the telescope's design, and optimizing it could result in more comprehensive time-series observations.

\begin{deluxetable*}{ccccc}
	\tablenum{3}
	\tablecaption{The fitting of planetary orbital parameters with simulated relative measurements.\label{Fitting}}
	\tablewidth{0pt}
	\tablehead{
		\colhead{Object} &\colhead{Parameters} 			& \colhead{Single planet fitting} 	&\colhead{Double planet fitting}		& \colhead{Given values}
	}
	\startdata
	Inner planet&$\cos i$      		& $0.853^{+0.013}_{-0.496}$  & $0.803^{+0.135}_{-0.138}$  & 0.867    			\\
	&$e \cos \omega$     		& $0.042^{+0.434}_{-0.003} $ &	$ 0.031^{+0.125}_{-0.161}$& 0.087     		\\
	&$e \sin \omega$     		& $0.000^{+0.662}_{-0.000}$  & $ 0.095^{+0.729}_{-0.084}$& 0.0500      		\\
	&$\Omega$ (rad)     	& $3.14^{+0.01}_{-2.31}$    & $1.52^{+1.08}_{-1.01}$& 1.05     		\\
	&$M_{0}$ (rad)       & $2.69^{+0.32}_{-0.16}$    & $1.54^{+1.06}_{-1.03}$& 0.79     		\\
	&Mass ($M_{\oplus}$) & $0.969^{+0.733}_{-0.016}$ &   $0.958^{+0.035}_{-0.033}$ & 1.000        			\\
	&Period (days)  		& $466.14^{+0.38}_{-102.30}$ & $453.89^{+17.63}_{-16.62}$& 447.67    		\\	
	Outer planet&$\cos i$     & 		& $0.730^{+0.047}_{-0.022}$     & 0.867  			\\
	&$e \cos \omega$     	&	& $0.070^{+0.017}_{-0.014} $	& 0.00   		\\
	&$e \sin \omega$     	&	& $ 0.171^{+0.013}_{-0.027}$   & 0.100      		\\
	&$\Omega$ (rad)     &	& $0.763^{+0.103}_{-0.049}$     & 0.785    		\\
	&$M_{0}$ (rad)      & & $0.383^{+0.056}_{-0.026}$     & 1.047    		\\
	&Mass ($M_{\oplus}$)& & $19.402^{+0.596}_{-0.299}$     & 20.000       			\\
	&Period (days)  	&	& $1229.51^{+37.42}_{-21.82}$ & 1266.21   		\\	
	\enddata
	\tablecomments{
	$i$ : inclination;	$\omega$ : argument of periapsis; $e$ : eccentricity; $\Omega$ : longitude of ascending node ;$M$ : mean anomaly; $M_{\oplus}$ : mass of the earth	
	}
\end{deluxetable*}

The targets are identified within this observing strategy, which also requires that observations alternate between these target stars. Consequently, this leads to a non-uniform distribution of observation times for the targets, further compounded by the fact that the observing strategy changes from year to year. While this variability enables CHES to encompass, to the greatest extent, the parallax period of the target stars and the period of planetary perturbations over the overall 5-year observation period, the resulting non-uniform sampling also impacts the fitting process. In the future study, additional techniques or algorithms are required to minimize errors and ensure micro-arcsecond precision measurements and planetary orbit detection.

\acknowledgments
We thank the referee for constructive comments and suggestions to improve the manuscript. This work is financially supported by the National Natural Science Foundation of China (grant Nos. 12033010 and 11773081), the Strategic Priority Research Program on Space Science of the Chinese Academy of Sciences (Grant No. XDA 15020800), and the Foundation of Minor Planets of the Purple Mountain Observatory.

\appendix
\section{Parameters related to the observed target}
In Section~\ref{sect:Prior}, we investigate the prioritization of observing targets, and the outcomes are elucidated in Table~\ref{table:appendix}. Ultimately, we carefully selected 94 target stars based on the criterion of the number of observations, with exclusion of those demanding an excessive number (more than 300). For the case where there are two target stars in the field of view (11 groups in total), they will be observed simultaneously, so the target star requiring more observations dominates in the prioritization calculation. The data columns in the table present five factors affecting priority: distance, V-band magnitude, number of reference stars, confirmed to have exoplanet or tentative candidates, and whether the star is part of a binary system. In the "Binary" column, the case of three stars is also included. We also present the spectral type of the target star in one of the columns. In the last two columns, we provide the number of observations and their priority.

\setcounter{table}{3}

\begin{longtable}{lccccccrr}
	\caption{Priority of target stars and number of observations\label{table:appendix}} \\
	\hline
	{Target stars} & {Distance} & {Magnitude} & {N$_{ref}$} &{Exoplanet}& {Binary}
	& {Spectral Type} &{N$_{obs}$}  & {Priority} \\
	 &{(pc)}&{V}& & {confirm}& &       & & {\%} \\
	\hline
	\endhead	
	\hline
	\endfoot
	\hline
	\endlastfoot
	 alf Cen A  & 1.35                             & 0.01                             & 59                           & N & Y                                  & G2V      & 30                              & 100.0                           \\
	 (alf Cen B) &&&&&&&&\\
	 omi02 Eri  & 5.04                             & 4.43                             & 16                            &Y& Y                                  & K0V      & 30                            & 99.89                           \\
	 alf CMi    & 3.51                             & 0.37                             & 31                            &N& Y                                  &  F5IV      & 30                            & 96.26                       \\	
	 eps Eri    & 3.20                             & 3.73                             & 16                            &Y & N                                    &  K2V      & 30 & 87.31                                                      \\
	 eta Cas A  & 5.84                             & 3.44                             & 38                            &N& Y                                 & F9V       & 30  & 86.10                                                     \\
	 (eta Cas B) &&&&&&&&\\
	
	 chi Dra    & 8.30                             & 3.58                             & 25                            &N& Y                                 &  F7V       & 57     & 80.72                                                 \\
	 61 Cyg B   & 3.49                             & 6.03                             & 46                            &N& Y                                  & K7V      & 78   & 78.20                                                     \\
	 (61 Cyg A) & & & & & & & &\\
	 36 Oph B   & 5.96                             & 5.03                             & 20                            &N& Y                                  & K1V      & 71 & 76.45                                                      \\
	 (36 Oph A) & & & & & & & &\\
	 70 Oph B   & 5.12                             & 6.07                             & 33                            &N& Y                                   & K4V      & 30 & 75.88                                                      \\
	 HD 191408    & 6.02                             & 5.32                             & 18                          &N & Y                                    & K2.5V      & 30& 75.71                                                      \\
	 HD 131977    & 5.88                             & 5.72                             & 14                           &N & N                                  & K4V   & 114&75.23                                                      \\
	 HD 219134    & 6.53                             & 5.57                             & 32                            &Y & N                                    & K3V    & 30&74.53                                                      \\
	 bet Hyi    & 5.84                             & 2.79                             & 20                            &N & N                                    & G0V      & 31 & 73.80                                                       \\
	HD 156384    & 6.84                             & 5.89                             & 13                            &N& Y                                  & K3V+K5V     & 42& 73.67                                                       \\
	 61 Vir     & 8.51                             & 4.74                             & 12                           &Y & N                                    & G7V      & 61 & 73.64                                                      \\
	 ksi Boo B  & 6.75                             & 6.82                             & 13                            &N& Y                                  & K5Ve      & 52 & 73.07                                                      \\
	 (ksi Boo A) & & & & & & & &\\
	 eps Ind    & 3.64                             & 4.69                             & 10                            &Y& N                                   &  K5V      & 31 & 72.78                                                      \\
	HD 102365    & 9.29                             & 4.88                             & 16                            &Y & N                                     & G2V      & 60& 72.08                                                       \\
	 p Eri B    & 8.19                             & 5.80                             & 17                          &N& Y                                 & K2V      & 119& 71.88                                                      \\
	 (p Eri A) &&&&&&&&\\
	 gam Vir B  & 12.7                            & 3.85                             & 13                         &N& Y                                  & F0mF2V      & 138 & 71.63                                                     \\
	 (gam Vir A) &&&&&&&&\\
	 41 Ara A   & 8.79                             & 5.52                             & 36                          &N& Y                                  & G9V      & 83  & 71.43                                                     \\
	tau Cet    & 3.60                             & 3.50                             & 7                             &Y & N                                    &  G8V    & 30& 71.20                                                      \\
	HD 192310    & 8.80                             & 5.72                             & 12                          &Y & N                                    & K2+V      & 107  & 71.12                                                    \\
	 iot Peg    & 11.8                            & 4.20                             & 27                            &N& Y                                 & F5V      & 51& 71.00                                                          \\
	V* AK Lep    & 8.90                             & 6.15                             & 21                            &N& Y                                  &  K3     & 119  & 70.52                                                    \\
	 (gam Lep)  &&&&&&&&\\
	HD 32450     & 8.38                             & 8.32                             & 14                            &N& Y                                   &  K7V      & 62 & 70.41                                                      \\
	 pi.03 Ori  & 8.04                             & 3.19                             & 19                            &N & N                                    & F6V      & 30 & 68.49                                                      \\
	 lam Ser    & 11.8                            & 5.03                             & 24                            &Y & N                                     & G0-V      & 119 & 68.06                                                     \\
	 zet TrA    & 12.2                            & 5.46                             & 29                            &N& Y                                  & F9V      & 102   & 66.73                                                   \\
	HD 85512     & 11.3                            & 8.83                             & 32                            &Y & N                                    & K6Vk      & 92  & 66.35                                                     \\
	 zet01 Ret  & 12.0                            & 6.18                             & 12                            &N& Y                                  & G2.5     & 155 & 66.10                                                      \\
	 (zet02 Ret) & & & & & & & &\\
	HD 38A       & 11.5                            & 10.2                            & 21                            &N& Y                                  & K6V      & 126 & 65.98                                                     \\
	85 Peg     & 12.7                            & 6.42                             & 15                            &N & Y                                    & G5V      & 130& 65.08                                                      \\
	HD 69830     & 12.6                            & 6.74                             & 24                            &Y& N                                    &  G8:V      & 160 & 65.02                                                     \\
	HD 40307     & 12.9                            & 8.10                             & 23                            &Y & N                                    &  K2.5V      & 121 & 64.13                                                     \\
	 e Eri      & 6.00                             & 4.27                             & 6                             &Y & N                                    &  G6V       & 49  & 63.42                                                    \\
	 bet TrA    & 12.4                            & 3.14                             & 53                            &N & N                                     &  F1V      & 121 & 63.21                                                     \\
	 sig Dra    & 5.77                             & 4.68                             & 27                            &N & N                                   & K0V       & 39    & 62.25                                                  \\
	HD 88230     & 4.87                             & 6.61                             & 16                           &N & N                                    & K6V      & 125 & 60.64                                                     \\
	 zet Tuc    & 8.53                             & 4.23                             & 15                            &N & N                                    & F9.5V      & 51 & 60.57                                                      \\
	 gam Pav    & 9.27                             & 4.22                             & 20                           &N & N                                   & F9V      & 60  & 59.73                                                    \\
	V* V2215 Oph & 5.95                             & 6.34                             & 21                            &N & N                                    & K5V    & 57 & 59.58                                                       \\
	 chi01 Ori  & 8.84                             & 4.40                             & 29                            &N & N                                    & G0V      & 60  & 59.56                                                     \\
	 mu. Cas    & 7.55                             & 5.17                             & 22                           &N & N                                     & G5Vb     & 33 & 59.07                                                       \\
	 107 Psc    & 7.61                             & 5.24                             & 12                           &N & N                                    & K1V      & 65  & 58.89                                                     \\
	 LHS 2713     & 11.0                            & 6.66                             & 3                             &Y & Y                                   & K2V      & 220   & 58.85                                                  \\
	 i Boo A    & 12.9                            & 5.72                             & 10                            &N& Y                                  & F5V     & 139   & 58.76                                                   \\
	 (i Boo B)	& & & & & & & & \\
	HD 147513    & 12.9                            & 6.02                             & 9                             &Y & N                                  & G5V       & 153  & 58.52                                                   \\
	V* TW PsA    & 7.61                             & 6.48                             & 15                           &N & N                                    & K4Ve      & 101& 57.65                                                      \\
	HD 157881    & 7.71                             & 7.56                             & 13                            &N & N                                    & K7V      & 39  & 57.17                                                     \\
	 gam Ser    & 11.3                            & 4.34                             & 17                            &N & N                                     & F6IV      & 59 & 56.92                                                      \\
	 61 UMa     & 9.58                             & 5.34                             & 12                            &N & N                                    & G8V      & 88 & 56.49                                                      \\
	HD 32147     & 8.85                             & 6.21                             & 20                            &N & N                                    & K3+V     & 175 & 56.43                                                     \\
	HD 50281     & 8.75                             & 6.57                             & 42                            &N & N                                    & K3.5V      & 50& 56.34                                                       \\
	 tet Per    & 11.1                            & 4.62                             & 23                            &N & N                                    & F8V      & 67 & 56.19                                                      \\
	 alf Men    & 10.2                            & 5.09                             & 25                            &N & N                                    &  G7V       & 81  & 56.18                                                    \\
	HD 100623    & 9.54                             & 5.98                             & 21                            &N & N                                   &  K0V      & 143& 55.82                                                      \\
		HD 10780     & 10.0                            & 5.63                             & 21                            &N& N                                  & K0V     & 97 & 55.59 														\\
	HD 17925     & 10.4                            & 6.05                             & 12                            &N & N                                    & V-K1V      & 145 & 54.85                                                     \\
	 zet Dor    & 11.6                            & 5.22                             & 30                            &N & N                                     & F9VFe-0.5     & 85   & 54.35                                                     \\
	HD 13445     & 10.8                            & 6.17                             & 6                             &Y & N                                     &  K1.5V    & 231  & 54.29                                                     \\
	HD 154363    & 10.5                            & 7.71                             & 18                            &N & N                                    & K4/5V      & 259  & 54.10                                                     \\
	 11 LMi     & 11.2                            & 6.11                             & 12                            &N & N                                     & G8V      & 109 & 53.86                                                     \\
	 bet Com    & 9.18                             & 4.25                             & 8                             &N & N                                     & F9.5V      & 70  & 53.74                                                     \\
	 54 Psc     & 11.1                            & 6.71                             & 19                            &Y& N                                    & K0.5V      & 168 & 53.62                                                     \\
	 lam Aur    & 12.5                            & 5.33                             & 27                            &N & N                                     & G1.5IV      & 70& 53.21                                                       \\
	V* V2689 Ori & 11.4                            & 10.3                            & 27                            &N & N                                   & K6V       & 166  & 52.87                                                   \\
	CD-57 1079   & 11.7                            & 10.4                            & 28                            &N & N                                     &  K7Vk     & 140  & 52.58                                                     \\
	HD 4628      & 7.4                             & 5.74                             & 9                             &N & N                                    & K2.5V      & 80  & 52.57                                                     \\
	HD 72673     & 12.2                            & 7.17                             & 32                            &N & N                                    & K1V      & 288 & 52.31                                                     \\
	HD 37394     & 12.3                            & 7.07                             & 18                            &N& N                                  &  K1     & 163  & 52.22                                                   \\
	 kap01 Cet  & 9.15                             & 4.85                             & 11                           &N & N                                    & G5V      & 74  & 52.11                                                     \\
	 HD 21531     & 12.5                            & 9.71                             & 7                           &N & Y                                     &  K5V     & 143 & 51.71                                                     \\
	HD 82106     & 12.8                            & 8.22                             & 18                            &N & N                                    & K3V      & 308 & 51.47                                                     \\
	HD 101581    & 12.8                            & 8.83                             & 12                            &N & N                                    & K4.5Vk      & 132& 51.42                                                      \\
	 ksi UMa B  & 8.73                             & 4.73                             & 5                            &N & Y                                  & G2V      & 82  & 51.34                                                     \\
	 (ksi UMa A) & & & & & & & &\\
	HD 36003     & 12.9                            & 8.78                             & 25                            &N & N                                     & K5V     & 150  & 51.26                                                     \\
	HD 217357    & 8.23                             & 7.87                             & 8                             &N & N                                     & K7+Vk      & 187   & 50.89                                                   \\
	HD 103095    & 9.18                             & 6.45                             & 9                             &N & N                                     & K1V     & 277 & 50.33                                                      \\
	 12 Oph     & 9.92                             & 5.77                             & 11                            &N & N                                    & K0V      & 124  & 50.03                                                    \\
	 bet Vir    & 11.1                            & 3.60                             & 7                             &N & N                                    & F9V      & 47& 48.69                                                       \\
	 bet CVn    & 8.61                             & 4.25                             & 7                             &N & N                                     &  G0V     & 60  & 48.32                                                      \\
	HD 222237    & 11.5                            & 8.09                             & 11                          &N & N                                     &  K3+V      & 101 & 47.66                                                     \\

\end{longtable}

\bibliography{ms}{}
\bibliographystyle{aasjournal}

\end{document}